\documentclass[iop,apj,twocolumn]{aastex63}  
\pdfoutput=1   
\newcommand{\megasaura}{M\textsc{eg}a\textsc{S}a\textsc{ura}}

\newcommand{\rcsohthree}{RCS-GA~032727$-$132609}

\newcommand{\jwst}{\textit{JWST}}
\newcommand{\hst}{\textit{HST}}
\newcommand{\etal}{et~al.}
\newcommand{\Msun}{M$_{\odot}$}

\newcommand{\Msol}{M$_{\odot}$}

\newcommand{\ebv}{$E(B - V)$}
\newcommand{\Hunits}{km~s$^{-1}$~Mpc$^{-1}$}
\newcommand{\cgsflux}{erg~s$^{-1}$~cm$^{-2}$}

\newcommand{\kms}{\hbox{km~s$^{-1}$}}
\newcommand{\cc}{\hbox{cm$^{-3}$}}
\newcommand{\peryr}{\hbox{yr$^{-1}$}}
\newcommand{\ciiidoublet}{[C~III] 1907, C~III] 1909~\AA}

\newcommand{\oiidoublet}{[O II]~3727, 3729~\AA}
\newcommand{\oiiidoublet}{[O III]~4959, 5007~\AA}
\newcommand{\oiiiuv}{[O III]~1660, 1666~\AA}

\newcommand{\hab}{H$\alpha$/H$\beta$}

\newcommand{\ArII}{\hbox{{\rm Ar}\kern 0.1em{\sc ii}}}
\newcommand{\ArIII}{\hbox{{\rm Ar}\kern 0.1em{\sc iii}}}
\newcommand{\CIV}{\hbox{{\rm C}\kern 0.1em{\sc iv}}}
\newcommand{\HI}{\hbox{{\rm H}\kern 0.1em{\sc i}}}
\newcommand{\HII}{\hbox{{\rm H}\kern 0.1em{\sc ii}}}
\newcommand{\HeI}{\hbox{{\rm He}\kern 0.1em{\sc i}}}
\newcommand{\HeII}{\hbox{{\rm He}\kern 0.1em{\sc ii}}}
\newcommand{\NII}{\hbox{{\rm N}\kern 0.1em{\sc ii}}}
\newcommand{\OI}{\hbox{{\rm O}\kern 0.1em{\sc i}}}
\newcommand{\OII}{\hbox{{\rm O}\kern 0.1em{\sc ii}}}
\newcommand{\OIII}{\hbox{{\rm O}\kern 0.1em{\sc iii}}}
\newcommand{\OIIlong}{{\rm O}\kern 0.1em{\sc ii}~$\lambda 3727$} 
\newcommand{\FeII}{\hbox{{\rm Fe}\kern 0.1em{\sc ii}}}
\newcommand{\NeII}{\hbox{{\rm Ne}\kern 0.1em{\sc ii}}}
\newcommand{\NeIII}{\hbox{{\rm Ne}\kern 0.1em{\sc iii}}}
\newcommand{\NeV}{\hbox{{\rm Ne}\kern 0.1em{\sc v}}}
\newcommand{\SII}{\hbox{{\rm S}\kern 0.1em{\sc ii}}}
\newcommand{\SIII}{\hbox{{\rm S}\kern 0.1em{\sc iii}}}
\newcommand{\SIV}{\hbox{{\rm S}\kern 0.1em{\sc iv}}}
\newcommand{\SiIV}{\hbox{{\rm Si}\kern 0.1em{\sc iv}}}
\newcommand{\MgII}{\hbox{{\rm Mg}\kern 0.1em{\sc ii}}}
\newcommand{\Halpha}{\hbox{{\rm H}\kern 0.1em$\alpha$}}
\newcommand{\Hbeta}{\hbox{{\rm H}\kern 0.1em$\beta$}}
\newcommand{\Heopta}{\hbox{{\rm He}\kern 0.1em{\sc i}}~$6678$}
\newcommand{\Heoptb}{\hbox{{\rm He}\kern 0.1em{\sc i}}~$5876$}
\newcommand{\Heoptc}{\hbox{{\rm He}\kern 0.1em{\sc i}}~$4471$}
\newcommand{\Brgam}{\hbox{{\rm Br}\kern 0.1em$\gamma$}}
\newcommand{\Brten}{\hbox{{\rm Br}\kern 0.1em$10$}}
\newcommand{\Breleven}{\hbox{{\rm Br}\kern 0.1em$11$}}
\newcommand{\HeIh}{\hbox{{\rm He}\kern 0.1em{\sc i}}~$1.7$~{\micron}}
\newcommand{\HeIk}{\hbox{{\rm He}\kern 0.1em{\sc i}}~$2.06$~{\micron}}
\newcommand{\squishlist}{
   \begin{list}{$\bullet$}
    { \setlength{\itemsep}{0pt}      \setlength{\parsep}{1pt}
      \setlength{\topsep}{3pt}       \setlength{\partopsep}{0pt}
      \setlength{\leftmargin}{1.5em} \setlength{\labelwidth}{1em}
      \setlength{\labelsep}{0.5em} } }
\newcommand{\squishend}{
    \end{list}  } 
\newcommand{\arcname}{SDSS~J1723$+$3411}
\newcommand{\linesdetected}{$42$}
\newcommand{\linesuplim}{$22$}
\newcommand{\bestz}{$z=1.3293$}
\newcommand{\logq}{\ensuremath{\log{(q)}}\,}
\newcommand{\lpok}{\ensuremath{\log{(P/k)}}\,}
\newcommand{\logOH}{12+$\log{\mathrm{(O/H)}}$\,}
\newcommand{\Ne}{n$_{e}$}
\newcommand{\Te}{T$_{e}$}

\shorttitle{Emission-line diagnostics in a lensed $z=1.32$ galaxy}
\shortauthors{Rigby \etal}
\received{19 June 2020}
\accepted{5 October 2020; ApJ in press}

\begin{document}
\title{A Comparison of Rest-frame Ultraviolet and Optical Emission-Line Diagnostics in the Lensed Galaxy \arcname\ at Redshift \bestz}
\author{J.~R.~Rigby}\affiliation{Observational Cosmology Lab, NASA Goddard Space Flight Center, Code 665, Greenbelt MD 20771, USA}
\author{Michael Florian}\altaffiliation{NASA Postdoctoral Program Fellow}\affiliation{Observational Cosmology Lab, NASA Goddard Space Flight Center, Code 665, Greenbelt MD 20771, USA}
\author{A.~Acharyya}\affiliation{The Australian National University, Australia}
\author{Matthew Bayliss}\affiliation{Department of Physics, University of Cincinnati, Cincinnati, OH 45221, USA}
\author{Michael D.~Gladders}\affiliation{Department of Astronomy \& Astrophysics, The University of Chicago, 5640 S.~Ellis Avenue, Chicago, IL 60637, USA}
\author{Keren Sharon}\affiliation{University of Michigan, Department of Astronomy, 1085 South University Avenue, Ann Arbor, MI 48109, USA}
\author{Gabriel Brammer}\affiliation{The Cosmic Dawn Center, University of Copenhagen, Denmark}
\author{Ivelina Momcheva}\affiliation{Space Telescope Science Institute, 3700 San Martin Dr., Baltimore, MD 21218, USA}
\author{Stephanie LaMassa}\affiliation{Space Telescope Science Institute, 3700 San Martin Dr., Baltimore, MD 21218, USA}
\author{Fuyan Bian}\affiliation{European Southern Observatory, Alonso de C\'ordova 3107, Casilla 19001, Vitacura, Santiago 19, Chile}
\author{H\r{a}kon Dahle}\affiliation{Institute of Theoretical Astrophysics, University of Oslo, P.O. Box 1029, Blindern, NO-0315 Oslo, Norway}
\author{Traci Johnson}\affiliation{Enterprise Analytics, CVS Health, 2211 Sanders Rd, Northbrook IL 60062, USA}
\author{Lisa Kewley}\affiliation{The Australian National University, Australia}
\author{Katherine Murray}\affiliation{Space Telescope Science Institute, 3700 San Martin Dr., Baltimore, MD 21218, USA}
\author{Katherine Whitaker}\affiliation{Department of Astronomy, University of Massachusetts Amherst, 710 N Pleasant Street, Amherst, MA 01003, USA}
\author{Eva Wuyts}\affiliation{ArmenTeKort, Antwerp, Belgium}
\email{Jane.Rigby@nasa.gov}

\begin{abstract} 
For the extremely bright lensed galaxy \arcname\ at \bestz ,
  we analyze spatially integrated MMT, Keck, and \textit{Hubble Space Telescope} spectra
  that fully cover the rest-frame wavelength range of 1400~\AA\ to 7200~\AA .
  We also analyze near-IR spectra from Gemini that cover H$\alpha$ for a portion of the lensed arc.
  We report fluxes for \linesdetected\ detected emission lines, and upper limits for an additional \linesuplim .
\added{This galaxy has extreme emission line ratios and high equivalent widths that are characteristic of extreme emission-line galaxies.}
  We compute strong emission line diagnostics from both the rest-frame optical and rest-frame ultraviolet (UV),
  to constrain physical conditions and test the spectral diagnostics themselves.
  We tightly determine the nebular physical conditions using the most reliable diagnostics, and then
  compare to results from other diagnostics.
  We find disappointing performance from the UV--only diagnostics: 
  \replaced{they are not able to measure the metallicity;}{they either are unable to measure the metallicity or
  dramatically under-estimate it;}  they over-estimate the pressure; and  
  the UV diagnostic of ionization parameter has a strong metallicity dependence in this regime.
  Based on these results, we suggest that upcoming \textit{James Webb Space Telescope} (\jwst) 
  spectroscopic surveys of galaxies in the reionization epoch 
  should invest the additional integration time to capture the optical [O~II] and [O~III] emission lines, 
  and not rely solely on the rest-frame UV emission lines.
  We make available the spectra; they represent one of the highest-quality emission
  line spectral atlases of star-forming galaxy available beyond the local universe, and will aid planning observations with \jwst .
\end{abstract}

\section{Introduction}  \label{sec:intro}
Nebular emission lines arise from the gas ionized by massive stars, and as such, 
are a fundamental way by which we understand the physical conditions within galaxies.  
Spectral diagnostics composed of the fluxes of two or more of these emission lines 
are used to measure nebular physical conditions, namely the gas pressure,
the ionization parameter (defined as the ratio of ionizing photons to electrons),
metallicity \replaced{(the ratio of oxygen to hydrogen atoms)}
{(expressed as the oxygen abundance relative to hydrogen, defined in units of $\log(O/H) + 12$) }, 
and the elemental abundance pattern.

While galaxies produce emission lines over much of the electromagnetic spectrum, 
the most heavily utilized spectral diagnostics have been in the rest-frame optical, 
in part because these diagnostics are accessible to ground-based telescopes 
\added{and have therefore been the workhorses of large, multiplexed spectroscopic surveys.} 
Indeed, the multi-object spectroscopy mode of the NIRSpec instrument on the 
upcoming \textit{James Webb Space Telescope} (\jwst) is designed to multiplexedly 
capture these rest-frame optical diagnostics 
in galaxies across much of cosmic time, \added{out to very high redshift.}

The rest-frame ultraviolet (UV) also features a number of emission lines \citep{Kinney:1993eh}
that are useful as spectral diagnostics \citep{Garnett:1997bo,Garnett:1997ec}.  
\added{Historically, sample sizes at low redshift have been limited due to the difficulty of detecting UV photons.}
Since the redder of these diagnostic emission lines will redshift out of NIRSpec's range 
for very high redshift galaxies, multiple groups have been calibrating rest--frame UV emission lines
as alternative diagnostics \citep{Bayliss:2014ib, Stark:2014fa, Berg:2016cj, Feltre:2016hg, Jaskot:2016wv, Steidel:2016jv,
Senchyna:2017if, Stark:2016fy, Stroe:2017hy, Stroe:2017gy, Berg:2018gd, Byler:2018jy, Nakajima:2018fg, 
Shibuya:2018ft, Kewley:2019kf, Berg:2019dr, Berg:2019gp, Acharyya:2019gg, Byler:2020gy}, 
\added{with small sample sizes.} 
These rest-frame UV line diagnostics are promising, but as yet have received
a small fraction of the observational and theoretical attention paid to the 
rest-frame optical lines.

The rest-frame optical diagnostics are either calibrated empirically from H~II regions at $z=0$, or from
theoretical models; these various calibrations disagree by up to a factor of five 
in the local universe \citep{Kewley:2008be}.  Furthermore, the rest-frame optical 
emission line ratios of galaxies have evolved systematically with time
\citep{Shapley:2005dl, Kriek:2007fw, Brinchmann:2008ie, Hainline:2009fg, Kewley:2013cs,
Steidel:2014es, Shapley:2015bd}.
The emission line ratios O32\footnote{O32 $\equiv$ [O III] 5007~\AA\ / [O II] 3727, 3729~\AA}
and [O~III]/H$\beta$, which are sensitive to ionization parameter,
are observed to rise from z$=$0 to z$=$0.6 \citep{Kewley:2015kr}, and are elevated in z$\sim$3 galaxies
\citep{Holden:2016iv, Onodera:2016aa}. One proposed explanation for this
observed redshift evolution in line ratios is that the ionization parameter evolves,
driven by higher specific star formation rates and/or star formation surface densities
\citep{Kewley:2015kr, Hirschmann:2017ch, Kaasinen:2018hva, Bian:2016jv}.

It is thus timely to compare these spectral diagnostics of physical conditions of 
galaxies \textit{in situ} at redshifts $z>1$, where diagnostic line ratios are known to be elevated.
It is especially important to obtain the full suite of diagnostics, both rest-frame UV and 
rest-frame optical, in order to understand any systematic biases.   Observationally this has
been difficult due to the faintness of typical galaxies at these redshifts, and due to 
the Earth's atmospheric opacity that prevents full wavelength coverage in the rest-frame optical 
(observed-frame near infrared).

A small number of papers have attempted joint analyses of the UV and optical emission lines
of galaxies.   \citet{Steidel:2016jv} compared the stacked rest-frame UV spectra of 
30 galaxies at $z \sim 2.4$ to their stacked rest-frame optical spectra, finding that the UV stellar 
continuum implied systematically lower metallicities, and attributing this to an atomic abundance
pattern with a much higher than solar ratio of alpha-process elements to iron-peak elements.
 The rest-frame UV emission lines were used to estimate
the oxygen abundance via the direct method, using the [O~III]~(1661 + 1666~\AA)/[O~III]~5007~\AA\ ratio.
\citet{Byler:2018jy} suggested the use of several other rest-frame UV emission lines as spectral
diagnostics; \citet{Byler:2020gy} tests the effectiveness of these diagnostics using
spectra of local and lensed $z\sim2$ galaxies.

The most in-depth intercomparison of the rest-frame UV and optical emission lines to date
has been carried out by \citet{Acharyya:2019gg}, hereafter A19; they jointly analyzed
the full suite of rest-frame UV and rest-frame optical emission lines in a single star-forming 
region within the $z=1.70$ lensed galaxy \rcsohthree, using spectra from 
\citet{Rigby:2011il} and \citet{Rigby:2018ev}.  A19 explored how what is inferred 
depends on what is measured---how the physical constraints depend on whether the
input spectra is the UV alone, the optical alone, or the full suite of rest-frame
UV and optical lines.  Using only the rest-frame UV lines, they were able to
constrain the ionization parameter but not the nebular metallicity.  They also inferred
systematically higher pressures from the rest-frame UV diagnostics compared to the
rest-frame optical diagnostics.

A limitation encountered by A19 was the difficulty of relative flux calibration (or ``fluxing'') 
across the broad wavelength range of the spectral diagnostics, for two reasons.
First, there was a wavelength coverage gap between the rest-frame UV and rest-optical spectra.  
Second, the near-IR spectra were obtained sequentially in the three filters (J, H, and K), as
the seeing and pointing varied.  While the Balmer line ratios were used to 
scale the relative fluxing across the three near-IR bands, 
relative fluxing across multiple spectra remained the main source of systematic error in that work.

Even at $z\sim0$, few galaxies have continuous spectral coverage of  
rest-frame 1000--7000~\AA, due to the difficulty of observing in the UV.  
Therefore, 1) to prepare for the sorts of datasets that \jwst\  
will soon obtain for galaxies at higher redshift, 
2) to evaluate the performance of the rest-frame UV and the rest-frame optical spectral diagnostics, 
and 3) to fully constrain the physical conditions of a distant star-forming galaxy, 
we have obtained new spectra for a lensed galaxy at \bestz\  
that completely cover this rest-frame wavelength range.

In this paper, for the bright gravitationally--lensed giant arc 
\arcname\ at \bestz\ \added{(Figure~\ref{fig:hstimage})}, we analyze spectra from instruments on four telescopes: 
the Blue Channel spectrograph on the MMT, 
the Echellette Spectrograph and Imager (ESI) on the Keck II telescope, 
the WFC3-IR G102 and G141 grisms onboard the \textit{Hubble Space Telescope} (\textit{HST}), 
and the Gemini Near-Infrared Spectrograph (GNIRS) on the Gemini-North telescope.
All but the Gemini/GNIRS spectra are spatially integrated over the giant arc;
the Gemini/GNIRS spectrum covers a small portion of the arc. 
\added{See Figure~\ref{fig:finder1723}}. 
We stitch together these spectra to completely cover all rest-frame wavelengths
from $1375$ to $7230$~\AA . 
We publish these spectra, and report fluxes for \linesdetected\ detected emission lines
and upper limits for \linesuplim\ more emission lines.  
We believe this to be the most comprehensive set of
of emission lines yet published for a galaxy beyond the local universe.
From these emission line fluxes, we compute 
strong emission line diagnostics from both the rest-frame optical and rest-frame UV,
in order to constrain the physical conditions of this galaxy, and to test the spectral diagnostics themselves.

\section{Methods}

All spectra were corrected for foreground reddening from the Milky Way galaxy,
using the value of  \ebv  $= 0.03415$ measured by \citet{Green:2015cf}.\footnote{Queried using 
the python interface at http://argonaut.skymaps.info}
All spectra have had the barycentric correction applied, and all wavelengths
are listed in vacuum.  
Rest wavelengths are from NIST.\footnote{\url{http://www.pa.uky.edu/\%7Epeter/atomic/}}  

\added{We use a solar oxygen abundance of  $12 + \log(O/H) = 8.72$ \citep{Asplund:2009eu}.}  

We use the convention that negative equivalent width indicates emission, and positive equivalent width indicates absorption.

To convert apparent magnitude to absolute, we assume a $\Lambda$CDM cosmology with 
$\Omega_M = 0.3$, $\Omega_\Lambda = 0.7$, and $h_0 = 70$~\Hunits . 

\subsection{Target and experimental design}
The galaxy \arcname, a gravitationally lensed giant arc at  \bestz, 
was independently discovered by multiple groups from 
Sloan Digital Sky Survey (SDSS) data. 
\citet{Kubo:2010kg} reported a redshift for the  lensing cluster of $z=0.4435 \pm 0.0002$,  and a
redshift for the giant arc of $z=1.3294 \pm 0.0002$, as measured with the
DIS spectrograph on the Apache Point 3.5~m telescope.
The CASSOWARY team \citep{Stark:2013fe} independently found the arc, 
reporting a lens redshift of $z=0.444$ and an arc redshift of $z=1.328$, 
from spectroscopy with the Blue Channel and Red Channel spectrographs
on the MMT.
The arc was also independently selected by the Sloan Giant Arcs Survey (SGAS1) (Gladders \etal\ in prep.)
It was first observed by our collaboration in the follow-up imaging program for SGAS candidate lenses 
at the 2.56~m Nordic Optical Telescope, where we obtained 2x300s g-band exposures with the MOSaic CAmera (MOSCA) 
on UT date 2010-03-13, in 0.77\arcsec\ seeing.
An updated cluster redshift of $z= 0.44227 \pm 0.00009$ was reported by \citet{Sharon:2020hc}, using the 
SDSS DR12 spectrum of the brightest cluster galaxy.  

We chose this lensed galaxy for this experiment because its redshift 
puts all the rest-frame optical spectral diagnostics within reach of the 
WFC3-IR grism spectroscopy mode \hst, 
and all of the rest-frame UV diagnostics within reach of
ground-based telescopes.   Further, this redshift places a bright emission line 
or pair of lines into each region of spectral overlap:
the \ciiidoublet\ doublet is covered by both the MMT and Keck spectra;
the \oiidoublet\ is covered by both the Keck spectrum and the WFC3-IR G102 grism;
and H$\beta$ is covered by both WFC3-IR grisms.
This spectral overlap enables the combined spectral dataset to be relatively flux calibrated across 
the full wavelength range.

\subsection{Lens model}
The lens model for \arcname\ is described in \citet{Sharon:2020hc}.  
The cluster lenses \arcname\ into a classic five-image lensing configuration, 
with two merging images straddling the critical curve to form a highly magnified giant arc 
southeast of the BCG, a counter image north of the BCG, and two radial images west of the BCG.
\citet{Sharon:2020hc} identify a second lensed source at $z=2.165$, with four images, 
which are also used to constrain the lens model.

\begin{figure*}[ht]
\begin{center}
\includegraphics[width=5in]{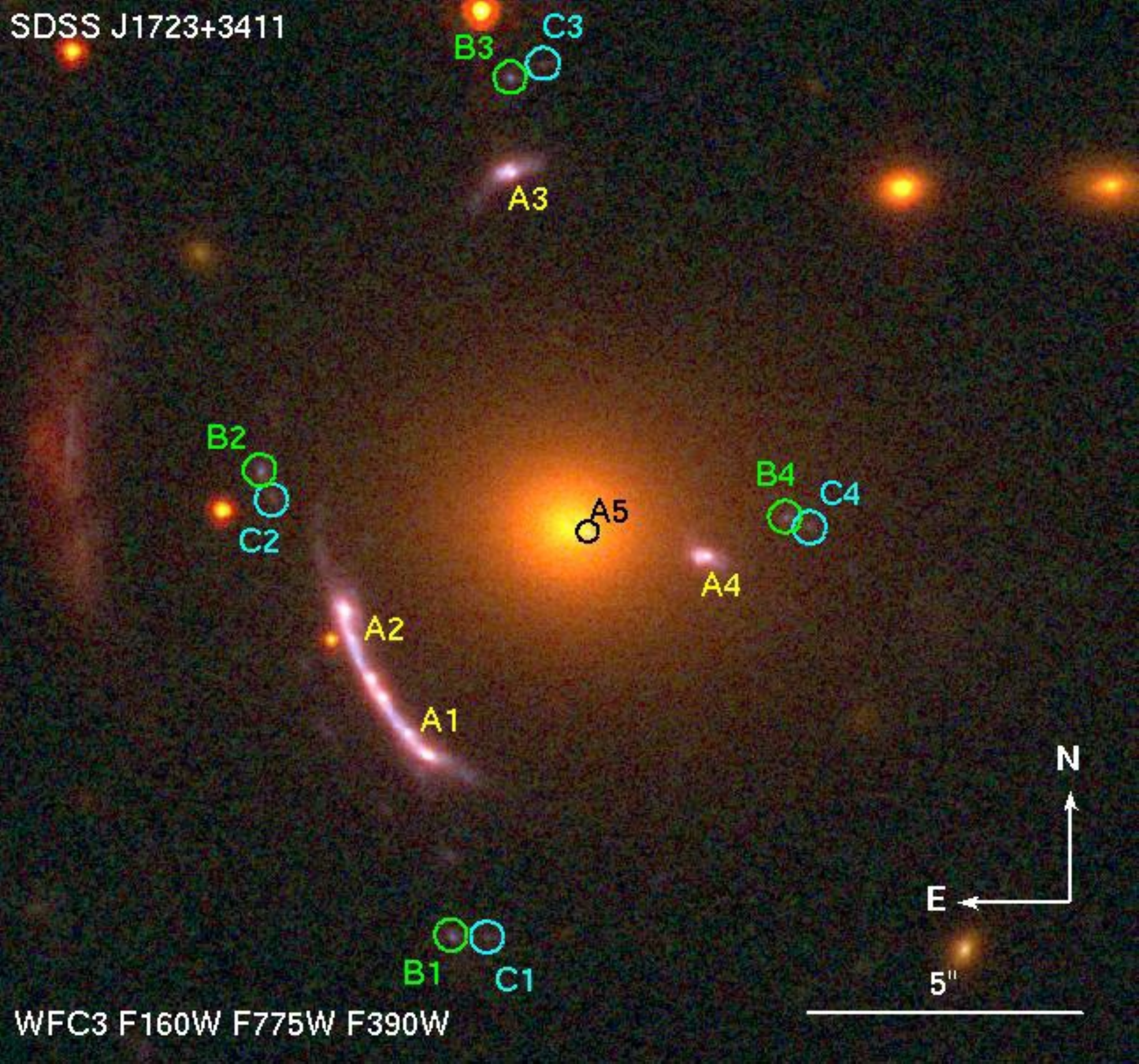}   
\figcaption{\textit{HST} color composite image of \arcname, with
RGB colors set to the  F160W, F775W, and F390W filters from the IR and UVIS and
channels of the WFC3 instrument.  Image families used in the lensing 
analysis of \citet{Sharon:2020hc} are marked;
the bright, multiply-imaged arc ``A,'' encompassing both
images A1 and A2,  is the subject of this paper.
\label{fig:hstimage}}
\end{center}
\end{figure*}

The lens model yields a total magnification for the giant arc of $52.7^{+3.3}_{-1.2}$ (Florian \etal\ submitted)
In general, uncertainty in lensing magnification propagates directly to absolute quantities like 
stellar mass and luminosity.  However, since gravitational lensing is achromatic, the magnification
uncertainty has no effect on quantities derived from flux ratios.  Thus, the 
physical conditions inferred later in this work are not affected by the lens model.
What is true is that certain portions of the arc are more highly magnified than others, and thus, 
are over-represented in the spatially-integrated spectrum analyzed in this paper.
This is an inevitable issue in lensing, that we must bear in mind when interpreting 
results from the integrated spectrum.  Florian \etal\ (submitted) analyzes the spatial variation in the
line ratios across physically distinct regions of the lensed galaxy, 
and discusses to what extent this affects the values inferred from integrated spectra.  
\added{To summarize, for \arcname, the line ratio with the most spatial variation (0.15 dex) is O32.  
R23 varies by 0.1 dex, and Ne3O2 does not vary significantly.  From MAPPINGS-V model grids, 
this spatial variation in line ratios can be explained by a spread of $\sim 0.25$ dex in ionization parameter, 
and a spread in metallicity of 0.3 dex (if on the high metallicity branch, which we think most likely) 
or zero dex (if on the low metallicity branch.) }

\begin{figure*}[ht!]
\begin{center}
\includegraphics[width=5in]{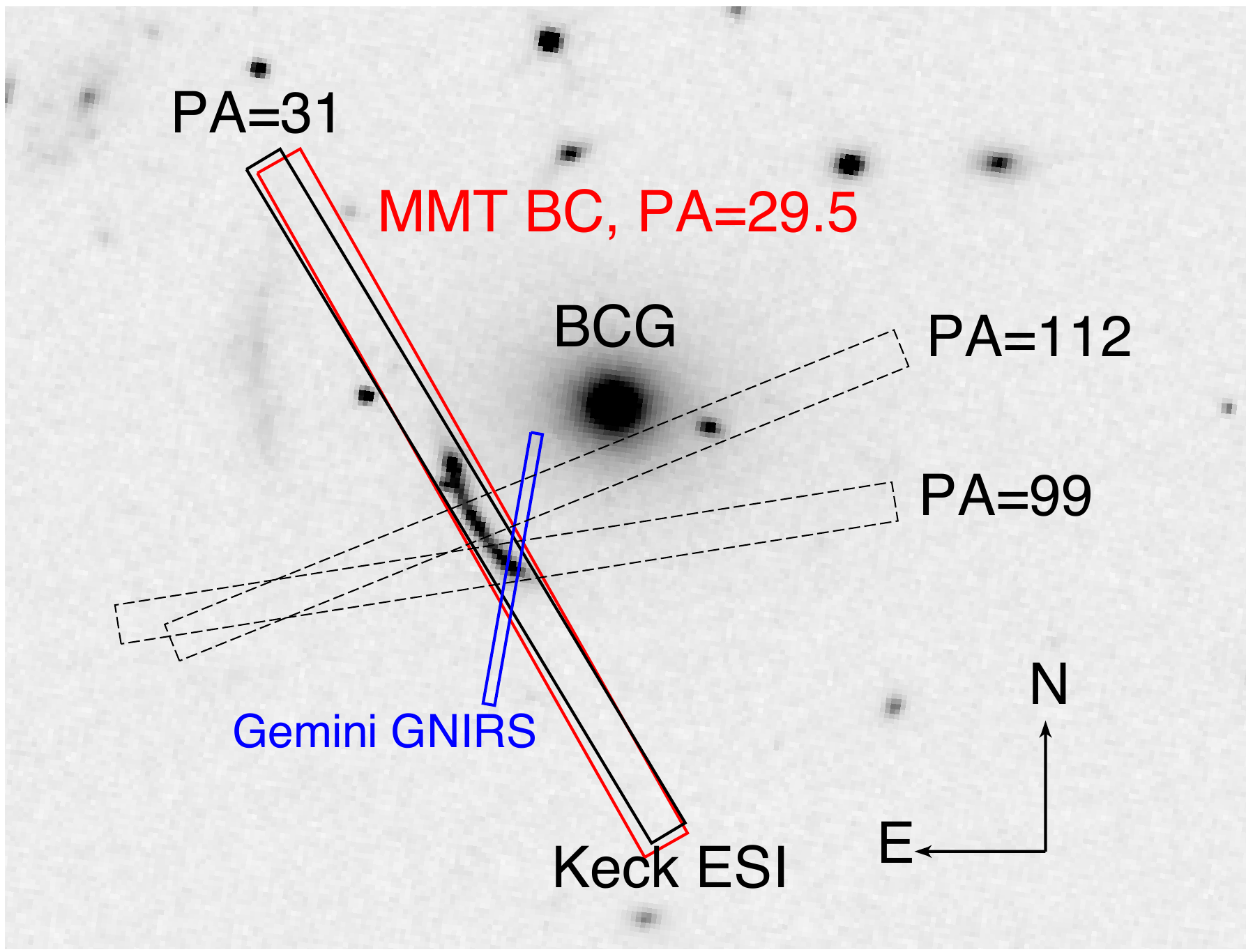}  
\figcaption{Finderchart to show spectroscopic slit positions for the
  lensed galaxy \arcname.  The background image is \textit{HST} WFC3-IR F105W.
  The long black rectangles show the Keck ESI pointings (with a 1 by
  20\arcsec\ long slit); the solid shape shows the pointing that
  captured the giant arc, which is used in this paper.  Two additional
  ESI pointings, which captured portions of the arc, are shown by the
  dashed rectangle.  The long red rectangle shows the MMT Blue Channel slit position
  (with a 1.25\arcsec\ longslit).  
The small blue rectangle shows the Gemini GNIRS pointing, with a 0.3\arcsec\ by 7\arcsec\ slit.  
The brightest cluster  galaxy (BCG) is marked.  
\label{fig:finder1723}}
\end{center}
\end{figure*}

\subsection{Observations and Data Reduction}

\subsubsection{\textit{HST} imaging data}
We obtained imaging of \arcname\ with the UVIS and IR channels of the WFC3 instrument
\citep{MacKenty:2012ky} onboard \hst\ through  guest observer program \#13003 (PI Gladders), 
which was a large program targeting lensed arcs selected by SGAS \citep{Sharon:2020hc} .
The filters used were F390W,  F775W,  F1110W, and F160W, at depths of
                      2368~s, 2380~s, 1112~s, and 1112~s respectively.
Figure~\ref{fig:hstimage} shows  a color composite \textit{HST} image of \arcname .
The giant arc of \arcname\ is labeled as
``A1'' and ``A2'' in Figure~\ref{fig:hstimage} and in \citet{Sharon:2020hc} .
The spectroscopy presented in this paper covers this giant arc. 
\textit{HST} grism spectra were obtained for additional lensed images of the galaxy,
labeled A3, and A4, that are not discussed in this paper.

Additional WFC3-IR imagery was also obtained in the F105W and F140W filters 
through guest observer program \#14230 (PI Rigby), as the non-dispersed images
used for wavelength calibration.

\subsubsection{\textit{HST} WFC3-IR grism spectra}
We obtained spectra of \arcname\ with the WFC3 instrument 
onboard \hst , using the IR channel and the 
G102 and G141 grisms, in guest observer program \#14230 (PI Rigby).
Eight orbits of grism spectroscopy were obtained on UT dates
2016-01-18,  2016-01-19, 2016-07-11, and  2016-07-14.
The total integration times were 2.76~hr in each of the G102 and G141 grisms.
Direct images for wavelength calibration were obtained using the F105W and F140W filters.
The grism observing strategy followed that developed for 3D-HST \citep{Brammer:2012bu}, 
namely the use of 2x2 interlacing and their four-point dither pattern.  
Observations were obtained at two  different roll angles (``ORIENT'' in the \hst\ nomenclature) 
of 139\arcdeg\ and 308\arcdeg .  From the range allowed by  spacecraft safety constraints,  
these orientations were chosen such that light was dispersed perpendicular to the length of the arc, 
while minimizing contamination from cluster galaxies.  
These two roll angles differ by 11\arcdeg\ (ignoring a meaningless 180\arcdeg\ rotation.)

\begin{figure*}
\begin{center}
  \includegraphics[width=5in,angle=0]{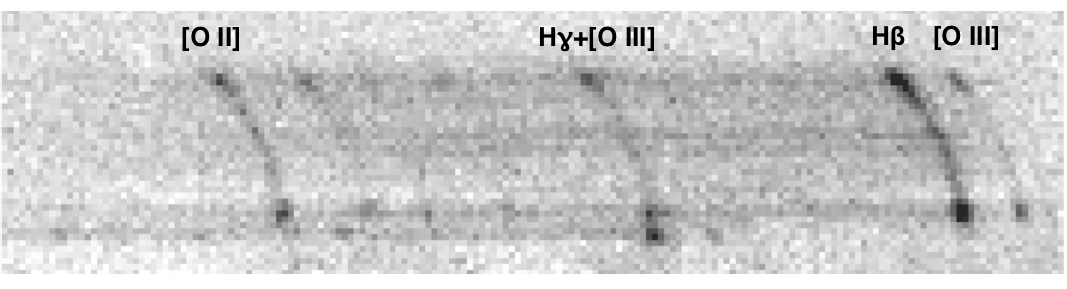}  
  
  \includegraphics[width=5in,angle=0]{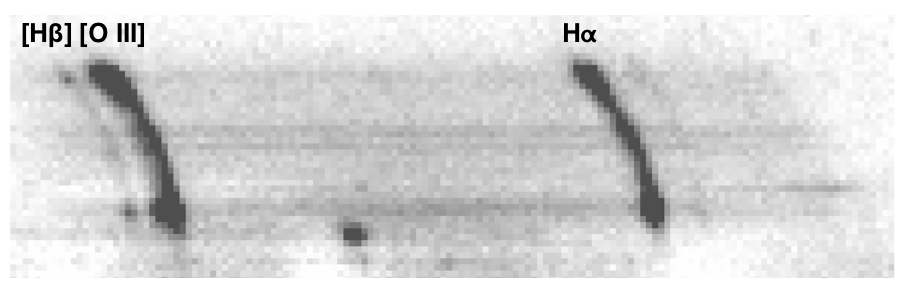}  
  \figcaption{Two-dimensional \hst\ WFC3-IR grism spectra of \arcname, at the 139\arcdeg\ roll angle,
    from the G102 grism (top panel) and the G141 grism (bottom panel).  
Emission from the sky and from contaminating neighbors has been removed.
The brightest emission lines are labelled.
\label{fig:grism2D}}
\end{center}
\end{figure*}

Grism data were reduced using the software package Grizli \footnote{https://github.com/gbrammer/grizli}.
We followed the steps of the standard Grizli reduction pipeline with a key modification.
Because this object is lensed by a galaxy cluster, there are several bright cluster members that
contribute significant contamination to the 2D grism spectrum; to best model the
contaminating light from these sources, their light profiles in the direct images were first modeled 
using GALFIT \citep{Peng:2010eh}.  Rather than assigning light to these objects using the Source
Extractor segmentation maps generated automatically by Grizli, light was assigned to these objects
based on their GALFIT models, with the light profiles truncated where the surface brightness fell to
less than $0.1\%$ 
of the central value.  These models were subtracted from the direct images, and
the preliminary contamination models for these objects were created separately from the
contamination models for the rest of the field.  After the initial models were made, the two sets of
preliminary contamination models were combined before running the standard contamination model
refinement steps in the Grizli pipeline.  When multiple objects contribute significant contaminating 
light to the same region in the observed 2D spectrum, this process can better assign that light to the
sources responsible for it, and therefore facilitate the production of better contamination models,
ultimately yielding cleaner 2D spectral extractions than those produced by the standard Grizli
procedure.  Figure~\ref{fig:grism2D} shows some of the 2D spectra after this process.

After extracting the 2D spectra from each grism and each roll angle, 1D spectra were produced by
using the flux-weighted center of the arc in each row of the direct images to determine the offsets in
the wavelength calibration from row to row in the 2D spectra.  We then interpolated each row to 
subsample the spectrum by a factor of 10, summing the rows in these finer wavelength bins, and 
then resampling the result to a coarser wavelength grid with bin sizes equal to the original dispersion 
(i.e., about 46\AA\ per pixel for G141, 23\AA\ for G102).
The extracted spectra cover the observed wavelength range of
$\sim 7800$--$11900$~\AA\ for G102 and 10400--17600~\AA\ for G141,
which corresponds to rest-frame wavelengths of 3350--5108~\AA\ and 4460--7555~\AA, respectively.
To maximize signal-to-noise ratios for
line-fitting, we summed together the 1D spectra from each of the two rolls;   
\added{Figure~\ref{fig:grismfits} shows these final spectra.}
Each 1D spectrum was also
fit individually as a consistency check; 
\replaced{the line ratios were found to be in agreement with
each other and with the summed spectrum within uncertainties.}
{the offset in the H$\alpha$/H$\beta$ ratio from roll to roll is considerably smaller
than the measurement uncertainty for a given roll.}

\subsubsection{MMT Blue Channel spectra}
We observed \arcname\ with the  Blue Channel spectrograph \citep{Angel:1979tv} 
on the 6.5~m MMT Observatory telescope on UT 2014-05-05  beginning at 09:20 UT.
Observing time was granted through the Harvard--Smithsonian Center for Astrophysics.  
Science exposures of 2400s and 1200s were taken at central wavelengths
of 4005\AA\ and 4205\AA, respectively, with the spectrograph
configured with the 1.25\arcsec\ wide longslit at a position angle
of 29.5\arcdeg\ East of North,  and the 800 line/mm grating in first order.  
The source was at low airmass, with $sec(z) < 1.03$. Conditions
were clear with  sub-arcsecond seeing during the science observations. 
The science frames were bracketed by HeNeAr arc lamp wavelength calibration frames and quartz
lamp flat calibration frames taken at the same position angle.
The spectrum was fluxed using observations of standard star Feige 34 that were taken the same night as the science observations,
with the same grating settings, and at the same airmass ($sec(z) = 1.02 \pm 0.02$).

The Blue Channel spectra were reduced as follows.  
The spectra were bias-subtracted, flat-field corrected, and wavelength
calibrated using standard IRAF routines from the
{\bf iraf.noao.imred.ccdred} and {\bf iraf.noao.onedspec} packages. We subtracted
the sky on a pixel-by-pixel basis using a two-dimensional model that
was generated with custom IDL code that makes use of the XIDL
package. The final object spectrum was boxcar extracted from the full
object profile of the giant arc, using an aperture that extended
approximately 3\arcsec\ along the spectroscopic slit. The final
combined spectrum covers a range in wavelength of
$\sim$3200--5200\AA, with a dispersion of 0.75\AA\ per pixel, spectral
resolution $R \equiv \delta \lambda / \lambda \simeq 1400$, and a median
signal-to-noise of $\sim 7$ per spectral pixel.

The Blue Channel  spectrograph is extremely sensitive in the blue, with sensitivity
down to 3000~\AA.  However, at 3000~\AA\ the Earth's atmospheric transmits only $0.6\%$, 
as compared to $17.7\%$ at 3200~\AA\ and $28.7\%$ at 3400~\AA\ (\citealt{Cox:2000ua} Table 11.25.)  
We therefore take $\lambda =3200$~\AA\ as the blue cutoff.
The spectrum effectively thus covers observed wavelengths of 3200--5200~\AA, which
correspond to rest-frame wavelengths of 1370--2230~\AA .

Due to a paucity of emission lines in the calibration lamps, the Blue Channel wavelength solution is 
uncertain by $\sim0.4$~\AA, as measured by comparing the centroids of bright sky lines to the 
lamp--based wavelength solution.

\subsubsection{Keck ESI spectra}
We obtained spectra of \arcname\ using the Echellette Spectrograph and Imager (ESI) 
on the Keck II telescope \citep{Sheinis:2002ft}, as follows.  
Observations were made on UT~2016-08-27 and UT~2016-08-28;
observing time was granted through Australian National University.
The echellette mode and the 1\arcsec\ slit were used for all observations, which provides
a spectral resolution of $R\approx 4000$.
Spectra were obtained at several different position angles,
as shown in Figure~\ref{fig:finder1723}.
For this paper, we use two integrations, with a total exposure
time of 2.055~hr, and a position angle (31\arcdeg\ E of N) that put the
entire arc  in the slit.  The airmass was $\sec (z) = 1.04$.
The spectrophotometric standard star BD$+$332642 was observed at the beginning of the night for
flux calibration.

We reduced the ESI spectra for each pointing using a combination of IRAF and python routines,
as follows.  First each raw 2D frame was bias and flat-field corrected.
Cosmic rays were removed using the IRAF routine \textit{lacos} \citep{vanDokkum:2001go}. 
Sky subtraction and spectral extraction were done using the IRAF tool \textit{apall}
within the {\bf noao.twodspec.apextract} package, fitting the trace with a second-order
Legendre polynomial, and using variance-weighted (also known as optimal) extraction
of the target spectra using a 33-pixel wide extraction aperture, and
with 17-pixel-wide sky apertures on either side of the target extraction aperture.
We performed the wavelength calibration to each extracted echelle order
by manually comparing the extracted spectra of CuAr, HgNe and Xe
arc lamp exposures to the corresponding template.
This led to a wavelength accuracy of at least $0.13\%$.
We used the \textit{standard}, \textit{sensfunc}, and \textit{calibrate} tasks in IRAF's
{\bf noao.onedspec} package to flux calibrate each extracted echelle order.
The same dispersion solution and flux calibration was applied to spectra from both nights.

We used custom python routines to combine the extracted echelle orders from each observation
of the science target within each night into one continuous, calibrated 1D spectrum,
using a mean weighted by the inverse variance in regions of order overlap.  We then summed the 1D spectra over
the multiple nights of observation, with inverse variance weights, applying the barycentric
correction in the process.
The resulting ESI spectrum covers observed wavelengths of 4000--10000~\AA, which corresponds to
rest-frame wavelength coverage of 1720--4290~\AA .

\subsubsection{Gemini GNIRS spectra}
We obtained spectra of \arcname\  with the Gemini Near-Infrared Spectrograph (GNIRS)
\citep{Elias:2006dh, Elias:2006jf} on the Gemini North telescope 
as part of program GN-2016B-FT-11 (PI Rigby).
Observations were obtained  on UT 2016-09-07, 
using the short camera, cross-dispersing prism (``SXD'' mode), 
the 0.3\arcsec\ slit, and the 111~lines/mm grating with the
central wavelength set to 1.529~$\mu$m.  This setup should provide a spectral resolution of $R=5900$.
The spectra covered observed wavelengths of 1.46--1.60~\micron\added{, which correspond to rest-frame wavelengths
of 0.627--0.687~\micron.}
Twelve integrations of 270~s duration were obtained; one suffered data quality issues and was discarded.
Thus, the effective integration time was 2970~s.
As shown in Figure~\ref{fig:finder1723}, the slit was centered on brightest knot in image A1, 
and was placed perpendicular to the giant arc.  As such, this spectrum covers a very different
portion of the giant arc than the Keck/ESI and MMT/Blue channel spectra.


The airmass ranged from $1.06 \le sec(z) \le 1.16$.
The data were obtained as A$-$B nods.
The grating position was not stable during the observations, which produced noticeable drift 
from frame to frame in the wavelength solution and in the location of the spectral trace on the detector.

\begin{figure*}[ht!]
\begin{center}
\includegraphics[width=5in,angle=0]{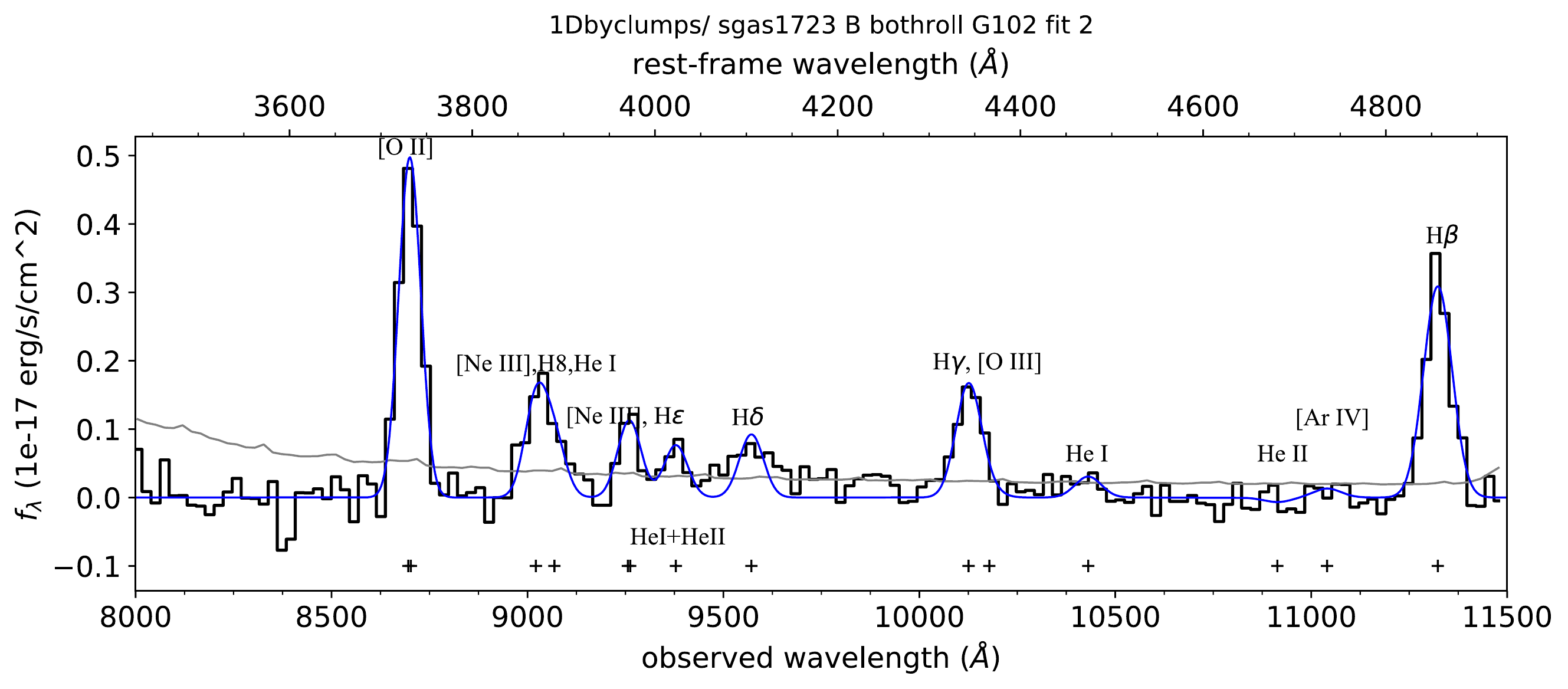}  

\includegraphics[width=5in,angle=0]{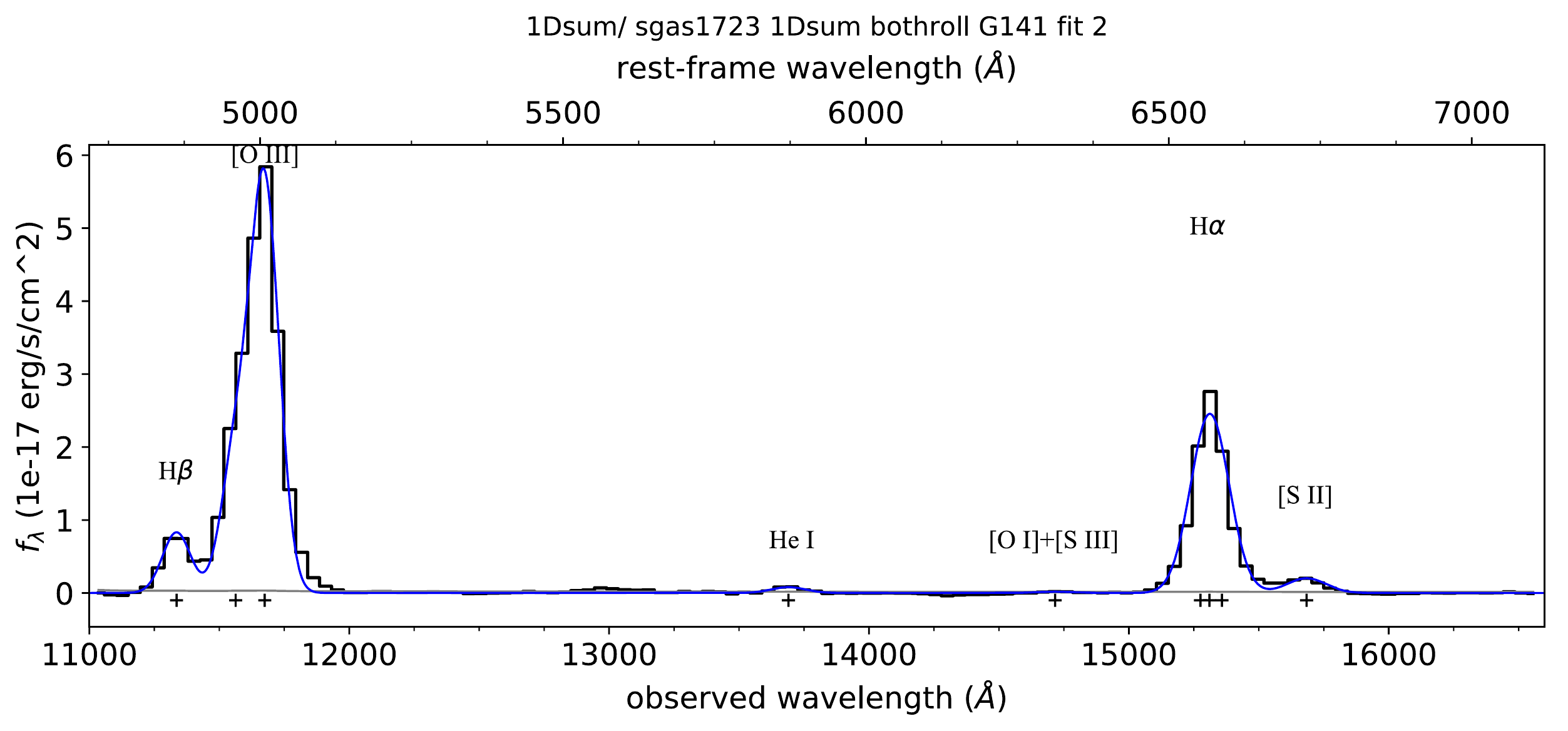}  
\figcaption{Line-fitting to the summed 1D \textit{HST} WFC3-IR grism spectra, for the 
G102 grism (top panel) and for the G141 grism (bottom panel.)  
The black thick line shows the continuum--subtracted spectrum,
the thin grey line shows the $1\sigma$ uncertainty spectrum, 
and the blue line shows the best fit.  Emission line centroids are marked with crosses.
\label{fig:grismfits}}
\end{center}
\end{figure*}

The GNIRS data were reduced to produce a one-dimensional spectrum as follows.
We used the Gemini IRAF package, with sky subtraction done by
differencing A$-$B pairs.  From each A$-$B pair, spectra of the A and B images were extracted
using a 7~pixel wide boxcar.   To mitigate the significant residual skylines due to the grating drift, 
we extracted a ``skyline residual spectrum'' from each A$-$B pair, 
using a 7~pixel wide boxcar located between the A and B positions of the galaxy.
We then subtracted the corresponding skyline residual spectrum from each individual
extracted spectrum.   The resulting 1D  spectra were then averaged, and the
error in the mean taken as the uncertainty spectrum.  The spectra were not absolutely fluxed.

\subsection{Relative flux corrections to the spectra}
The five spectra of \arcname --- from the MMT Blue Channel, Keck ESI, \hst\ WFC3-IR G102 and G141, and Gemini GNIRS ---
together provide overlapping, continuous wavelength coverage.  
The HST WFC3 grism spectra should be the best fluxed, since they 
do not suffer from seeing, atmospheric transparency variation, or slit losses.  
We therefore adjust the flux scales of all other spectra to match that of the G141
grism spectrum, as follows:

\begin{itemize}
\item We do not apply any relative flux scaling between the G102 and G141 grism spectra, 
since both are slitless spectra from a space telescope, and therefore do not suffer seeing--dependent
slit losses.  The flux calibration of the WFC2-IR grisms is excellent
(Kuntschner et al.\ 2011, ST-ECF ISR WFC3-2011-05) and has been temporally stable to better than $1\%$
over four years (Lee, Pirzkal, and Hilbert 2014, STScI ISR WFC3 2014-01).
 
\item We scale the flux of the ESI spectrum to match that of G102,
by using the \oiidoublet\ doublet.  The doublet is spectrally resolved
in the ESI spectrum but is blended in G102.  Therefore, we scale the flux
of the ESI spectrum so that the combined flux in the [O~II] doublet matches
that in the G102 spectrum. 
With this scaling, the continua of the ESI and G102 spectra overlap at 8470~\AA, 
close to the observed position of the \oiidoublet .

\item We scale the flux of the Blue Channel spectrum to that of the ESI 
spectrum, using the \ciiidoublet\ doublet.  The doublet
is spectrally well-resolved in the ESI spectrum, but is blended in the
lower--resolution Blue Channel spectrum.  Therefore, we scale the flux
of the Blue Channel spectrum so that the combined flux in the C~III doublet
matches that of the ESI spectrum.
With this scaling applied,  the  median $f_{\lambda}$ in the 
line--free region 4200--4300~\AA\ for Blue Channel and for ESI agree to within $2\%$.

\item The Gemini GNIRS spectra are not \added{absolutely} fluxed, so we do not attempt any relative flux
calibration.  None is needed, since the GNIRS spectrum is only used to obtain the flux ratio of the
[N~II] lines to nearby H$\alpha$.
\end{itemize}

\subsection{Measuring emission lines}
We measure emission line fluxes as follows; measurements are tabulated in Table~\ref{tab:linefluxes}.

\begin{figure}
\includegraphics[width=3.5in,angle=0]{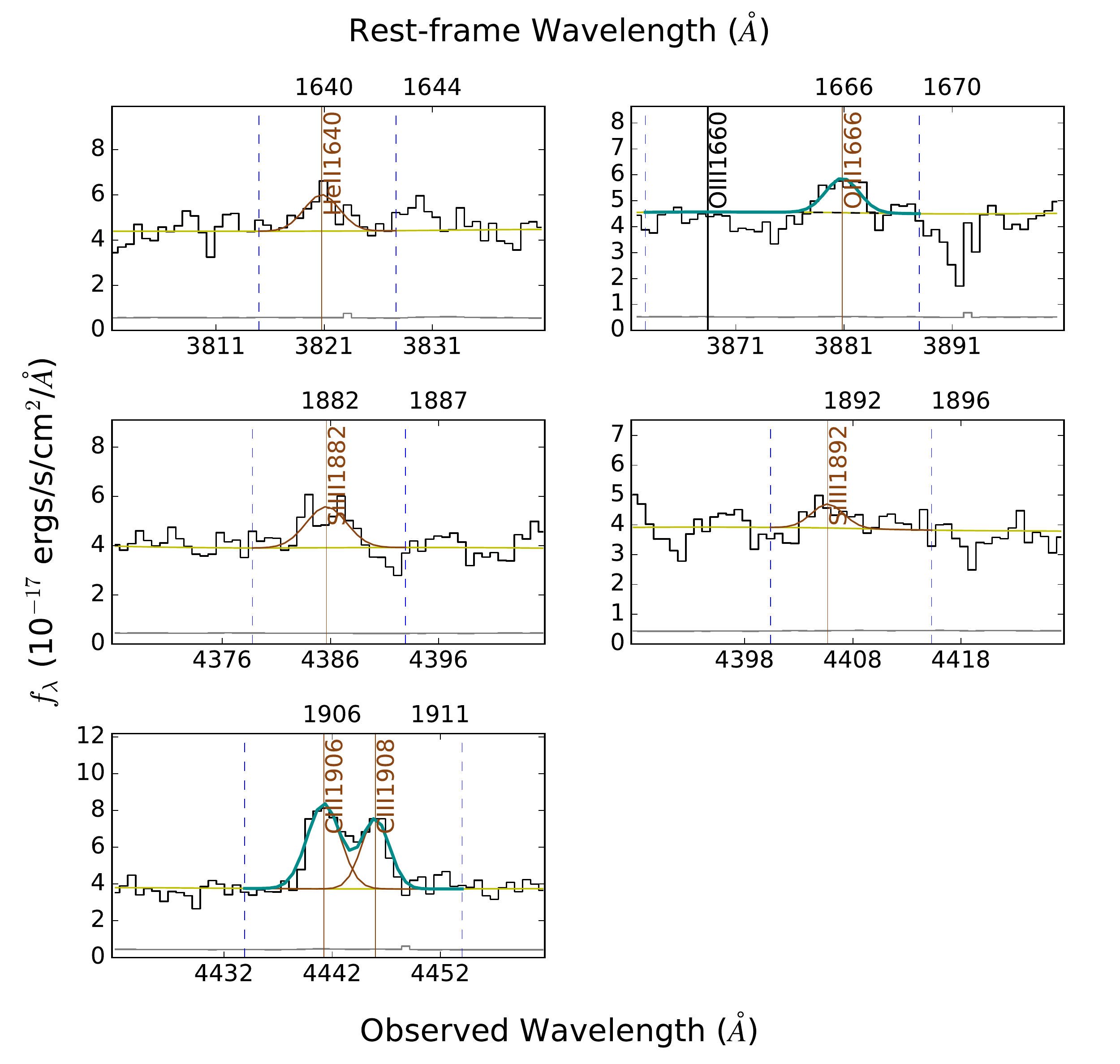}  
\figcaption{\added{Detected emission lines} and Gaussian line fits  in the MMT Blue
Channel spectrum of \arcname .  Black steps show the spectrum,  
\replaced{blue and red}{brown} vertical lines show the \deleted{expected and} 
best--fit central wavelength of  detected ($>3 \sigma$ ) lines, 
\replaced{red}{brown} Gaussians show the individual components of the fit, and thick green lines 
show the sum of multiple fitting components.  The $1\sigma$ uncertainty
spectrum is shown in grey. Black vertical lines denote non-detection ($<3 \sigma$). 
The fitting routine works on a portion of the spectrum as shown bounded by blue dashed lines.  
\deleted{Lines are ordered by increasing wavelength.}
\label{fig:MMTlines}}
\end{figure}

\subsubsection{Fitting lines in the ESI and MMT spectra}\label{sec:fitkeckmmt}
To fit an empirical continuum to each of of the MMT Blue Channel and Keck ESI spectra, 
we mask the expected positions of ISM absorption lines, nebular emission lines, and stellar wind lines, 
and then boxcar smooth the spectrum.

\begin{figure}
\includegraphics[width=3.5in,angle=0]{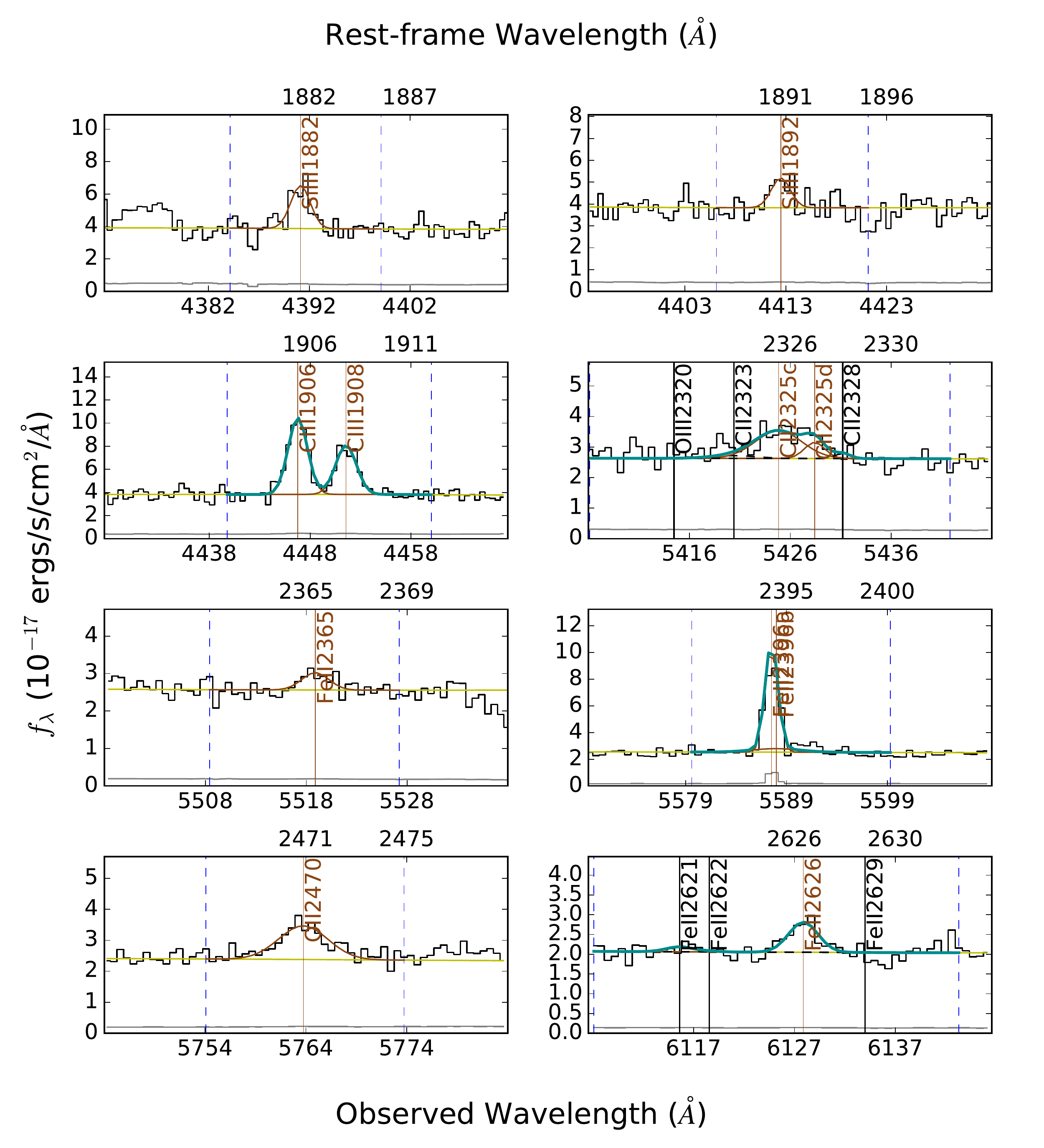}  

\includegraphics[width=3.5in,angle=0]{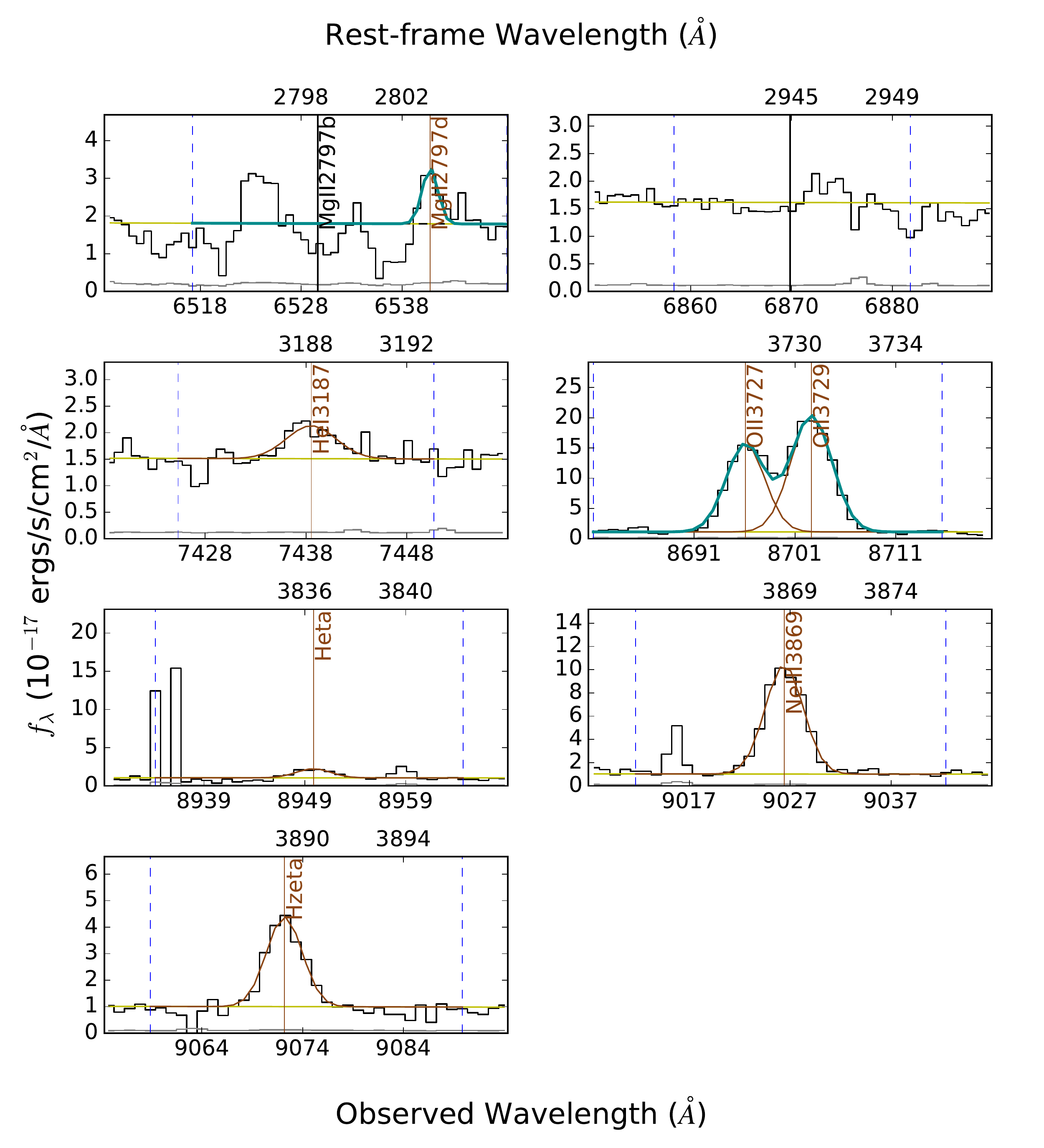}  
\figcaption{Detected emission lines and Gaussian line fits 
in the Keck ESI spectrum of \arcname .  Labels and color codes are as in 
Figure~\ref{fig:MMTlines}.
\label{fig:ESIlines}}
\end{figure}

For the continuum-subtracted Keck and MMT spectra, we simultaneously fit all emission 
lines using the methodology of A19.  Examples are shown in Figure~\ref{fig:MMTlines} and
Figure~\ref{fig:ESIlines}.
Briefly, neighboring emission lines (defined as within $\pm 5$ spectral resolution elements)
are fit with Gaussians simultaneously, with the line centers allowed to move by 
$3\sigma$ from the expected position given the systemic redshift, 
which we assume to be the redshift of H$\alpha$ measured in GNIRS/Gemini (see \S\ref{sec:halphaz}.)  
Line widths are bounded between a spectral resolution element and $300$~\kms .
\replaced{Significance of flux and equivalent width measurements are calculated following 
\citet{Schneider:1993hx}.  We consider lines to be detected if they
are $>3\sigma$ significant, and have measured flux greater than the flux uncertainty.
Features not satisfying these criteria are considered undetected; for these 
we quote $3\sigma$ upper limits.} 
{Flux and equivalent widths are measured from this fitting process.  We also report 
the significance of the feature following the method of \citet{Schneider:1993hx}, which 
indicates how significant the line is given the noise in the spectrum, without 
prior knowledge of the redshift.  We consider a line to be detected if has a Schneider \etal\ 
significance greater than 3, and if its measured flux is greater than the flux uncertainty.
For undetected features, we quote the Schneider \etal\ $3\sigma$ limits on flux and equivalent width.}

\subsubsection{Fitting lines in the WFC3-IR grism spectra}
\replaced{The very low spectral resolution of the grism spectra (see Figure~\ref{fig:grism2D})
requires a different approach to measuring emission line fluxes.}
{The grism spectra require a different approach to measuring emission line fluxes, since
the very low spectral resolution means that linewidths are set entirely by the disperser and by the
morphology of the target relative to the dispersion direction, and since the wavelength calibration
accuracy is much worse than from an echelle spectrograph.}
For each grism spectrum, we fit an empirical continuum using the XIDL package
x\_continuum, in which the portions of the spectrum that are free of emission lines 
were fit with a ninth-order Legendre polynomial. 
The emission lines in each continuum-subtracted grism spectrum were then fit as follows,
with separate fits to the G102 and G141 spectra.
We use the python package LMFIT, which implements non-linear least-square 
minimization and curve-fitting \citep{Newville:2016up}. 
The width of each emission line was not free to vary, but instead set by the following assumptions: 
that the instrumental line spread function is Gaussian, 
that the lines are spectrally unresolved, 
that the intrinsic spectral resolution $R_i$ of each grism (as would be appropriate for a point source)  
is $R_i=210$ for G102 and $R_i=130$ for G141 \citep{WFC3_DHv4.0}, 
and that the observed spectral resolution can be approximated as $R_o = R_i /m$, 
where $m$ is a morphological broadening factor for that spectrum, 
which is solved for in the fitting process.   
This morphological broadening occurs because the grism spectra are slitless;
it approaches unity for a thin source aligned perpendicular to the dispersion direction, 
and is high when the source elongation is aligned with the dispersion direction.

We fit 14 emission lines to the G102 spectrum, and 9 emission lines to the G141 spectrum, 
as shown in Figure~\ref{fig:grismfits}, with results listed in Table~\ref{tab:linefluxes}.  
Since the  [S~II] 6718, 6732~\AA\ doublet is unresolved at the grism resolution, and 
since the ratio is not constrained by other spectra, we fit it with a single Gaussian. 
We do the same for the blend of H8 and He~I near 3890~\AA, and the blend of [O~I] and [S~III] near 6310~\AA. 

We fix the following doublet ratios to their theoretical values from \citet{Storey:2000jd}:
[Ne~III] 3870.16, 3968.91~\AA; 
[N~II]  6549.85, 6585.28~\AA; and 
[O~III] 4960.295, 5008.240~\AA. 
At the low spectral resolution of the WFC3-IR grisms, the [N~II] doublet is blended with H$\alpha$,
and the \oiidoublet\ doublet is not resolved.  Therefore, we  
fix the [O~II] doublet ratio to that measured from the ESI spectrum (see \S\ref{sec:fitkeckmmt}), 
and fix the ratio of [N~II] 6585 / H$\alpha$ to that measured from the Gemini GNIRS spectrum (see \S\ref{sec:fitgnirs}).

The dispersion solutions for the WFC3-IR grisms have an internal accuracy 6\AA\ for G102 and 9~\AA\ for G141
(\citealt{WFC3_DHv4.0}, see \S9.3.8);  
these are significant fractions of a pixel.  
To compensate, we fit the spectra with a two-step process.
In the first step, we fit each spectrum by solving for a common redshift, a morphological 
broadening factor, and the flux of each unconstrained emission line.
In the second step, we fix the redshift as the result from step 1, and 
re-fit the spectra, again solving for the 
morphological broadening factor and the line fluxes, but this time allowing
the wavelengths of groups of neighboring lines to shift by a common offset.  
The resulting wavelength offsets are within the range expected given the 
internal accuracy of the wavelength solution.  

The grism spectra should have excellent fluxing accuracy, since they are slitless 
and obtained from space.  Small changes in the fluxing can occur due to roll-dependent differences
in the contamination model and the extraction region. 
For G102, the fluxes of the three brightest lines scale by  $2\%$ and $8\%$ for the 
two individual rolls compared to the spectrum extracted from both rolls combined.
For G141,  the fluxes of the three brightest lines scale by $-6\%$ and $+4\%$ for the 
two individual rolls compared to the spectrum extracted from both rolls combined.  
For the rest of the analysis, we use the fluxes extracted from the combined spectra from both rolls.  

In order to fit the multiple line blends in the grism spectra, it was necessary to assume
a line-spread function. We now examine the impact of that assumption, using a relatively isolated
line.   In G102, the brightest isolated line is H$\beta$: direct summation of the
continuum--subtracted spectrum returns $8\%$ higher line flux than from line fitting.
Direct summation produces fluxes that are $6\%$ higher for \oiidoublet\ (again in the G102 grism), 
and $3\%$ lower for the [N~II]+H$\alpha$ blend (in the G141 grism).   
Thus, there appears to be a several percent uncertainty in measuring a line flux, which may be attributed 
to the detailed shape of the line-spread profile.

\subsubsection{Fitting lines in the Gemini GNIRS spectrum}\label{sec:fitgnirs}
We fit the 1D GNIRS spectrum with Gaussians using the IDL Levenberg-Marquardt 
least-squares fitting code MPFIT \citep{Markwardt:2009wq}, 
following the method described in \citet{Wuyts:2014eu}.
The continuum was assumed to be flat over the wavelength region of interest, and the 
continuum level was allowed to vary.
The flux ratio of the [N~II] doublet was locked at the value from \citet{Storey:2000jd}. 
The central wavelengths of each line was set using the rest wavelength from NIST and 
the measured H$\alpha$ redshift in the 1D spectrum, and allowed to vary by $\pm3\times$ the
$1\sigma$ uncertainty in the central wavelength of H$\alpha$.  
The uncertainty spectrum was used as weights in the fitting, which prevented  
adjacent residual skylines from biasing the fit.  
The redder transition of [N~II] is adjacent to a skyline, but is clearly detected.
The linewidths were tied together, and the width of the linespread function was varied,
minimizing chi-squared.  

The [S II] doublet was not detected in the GNIRS spectrum, and so was not fit.

\begin{figure*}[ht!]
\includegraphics[width=7.2in]{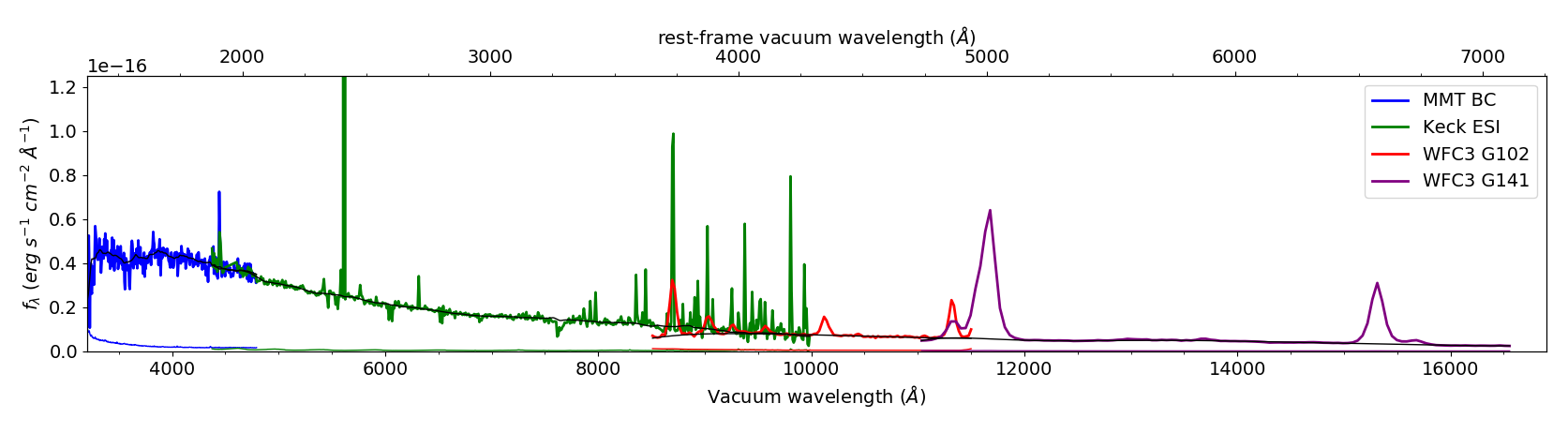}  
\figcaption{Four of the five spectra for \arcname .
From left to right, the spectra shown are from the MMT Blue Channel, Keck ESI, 
HST WFC3-IR G102, and HST WFC3-IR G141.  The MMT and Keck spectra have been 
median smoothed with a boxcar for readability.  
The flux scaling of the MMT and Keck spectra have been adjusted
as follows:  The Keck ESI spectrum has been scaled to match the flux in 
the \oiidoublet\ emission lines in the G102 spectrum, 
and the MMT Blue Channel spectrum has been scaled to match the flux 
in the \ciiidoublet\ emission lines in the ESI spectrum.
For readability, most of the overlap between the spectra is not shown.
The $1\sigma$ uncertainty spectra are shown with the same color-coding as the 
science spectra, and the fitted continua are shown as thin black lines.
\label{fig:prettyspec}}
\end{figure*}

\section{Results}\label{sec:results}

\subsection{A new spectral template for emission--line galaxies} 
In Figure~\ref{fig:prettyspec}, we present a contiguous spectral template of 
the bright gravitationally--lensed galaxy \arcname, 
completely covering the rest-frame wavelength range $1375$ to $7230$~\AA,
in which \linesdetected\ individual emission lines are detected. 
We electronically publish this template
and provide electronic versions of the reduced spectra from the four individual spectrographs.
We expect these spectra should be helpful to investigators preparing observing programs
with the \textit{James Webb Space Telescope}, as well as other applications.  

\subsection{Luminosity and stellar mass}
From the \hst\ imagery, Florian \etal\ (submitted) compute observed AB magnitudes for the giant arc of 
20.6,  20.5,  19.9,    20.2, 20.2,  and 20.1 in the F390W, F775W,  F105W, F110W, F140W, and F160W 
filters, respectively (see their Table~4.)  Given those magnitudes and the magnification 
reported by Florian et al., we calculate that the \deleted{the} absolute magnitude at 
rest-frame 1700~\AA\ is $-19.95$. 
This is very similar to the $M^{*}_{1700} = -19.8^{+0.32}_{-0.26}$ reported at $z=1.7$ by 
\citet{Sawicki:2006jva}.  As such, while \arcname\ appears to be one of the brightest lensed 
galaxies known, its intrinsic UV luminosity is typical for star-forming galaxies at its redshift. 

From H$\alpha$ fluxes measured in the \hst\ grism spectra, \added{and the conversion of \citet{Kennicutt:1998ki},} 
Florian \etal\ estimate a total 
star formation rate for \arcname\ of $7.9 \pm 0.4$~\Msun~\peryr, from the northern complete
image (image 3.)
Based on photometry from \hst\ and \textit{Spitzer}, \deleted{and} the lens model of \citet{Sharon:2020hc}, 
\added{and the Prospector MCMC-based stellar population synthesis code \citep{Johnson:2017bq},}
Florian \etal\ (submitted, see their Appendix) estimate a stellar mass for \arcname\ of 
$5.95^{+2.2}_{-1.86} \times 10^{8}$ \Msol .
Thus, \arcname\ lies a factor of 12 above the star formation rate--stellar mass relation at 
$1.0<z<1.5$ \citep{Whitaker:2014koa}.  
\replaced{Therefore, it is forming stars at a higher rate than is typical for a galaxy of its stellar mass.} 
{Comparisons to similar compilations by \citet{Tomczak:2016gg} and \citet{Santini:2017fy} yield
a similar result: \arcname\ lies about a factor of 10 above the  star formation rate--stellar mass relation 
for its redshift.  It is thus a ''starburst'' galaxy experiencing significantly more star formation than
is typical for galaxies of its stellar mass at $z=1.3$.}

\subsection{Redshift of H$\alpha$}\label{sec:halphaz}
To mitigate the problem that the grating drifted during the GNIRS observations,
we measure the redshift of $H\alpha$ from a subset of the two-dimensional GNIRS data, as follows.
For each integration, we produce an A$-$B (or B$-$A) difference image.  For the 6 cleanest
difference images (of 11), we measured the position of H$\alpha$ relative to the positions of 
two bright skylines (at known vacuum wavelengths of $\lambda_{vac} = 15241.0$ and  $15332.4$~\AA),
and assumed a linear dispersion across this small wavelength range.  
We test the accuracy and precision of this procedure by measuring the wavelength of a third,
fainter skyline, at $\lambda_{vac} = 15287.8$~\AA ; we recover its wavelength
to an accuracy of 0.1~\AA. 

We measure the wavelength of H$\alpha$,  and apply the barycentric correction,
resulting in a 
$\lambda_{vac}(H\alpha) = 15290.8 \pm 0.6$~\AA\ (quoting the median of the measurements
and the standard deviation.)
We thus measure the redshift as $z(H\alpha) =  1.3293 \pm 0.0002$.
This is the best determination of the systemic (nebular) redshift for this galaxy.

This measured redshift is fully consistent with that measured by
\citet{Kubo:2010kg}, of  $z=1.3294  \pm 0.0002$, measured 
from the \ciiidoublet\ and \oiidoublet .
This measured redshift is also similar to the redshift of 
$z=1.328$ (no uncertainty quoted) measured by \citet{Stark:2013fe} 
from the \ciiidoublet\ doublet and ISM absorption lines.

\subsection{Reddening}  
We use the Balmer line ratios to determine the amount of reddening of the nebular lines.
Table~\ref{tab:balmerratios} lists the observed line ratios and the inferred \ebv\ reddening 
for the spectra from each of the two roll angles, as well as from the roll--combined spectra.
By experimental design, H$\beta$ falls in the few hundred Angstrom region where the G102 and G141 grisms
overlap in wavelength at high sensitivity.
We believe the flux is measured more accurately and precisely by the G102 grism, for several reasons.  
First, the higher spectral
resolution of G102 more cleanly separates the H$\beta$ line from the [O~III] doublet.
Second, H$\beta$ in \arcname\ falls near the edge of the G141 wavelength coverage, 
where the flux calibration is more uncertain.
Third, in the 139\arcdeg\ roll, the G141 grism spectrum suffers from some contamination by the 
counter-image ``A4''  that affects the measurement of H$\beta$ .
The flux of H$\beta$ measured in each grism agrees to within $9\%$ (139\arcdeg\ roll) and $19\%$ (308\arcdeg\ roll). 
Table~\ref{tab:balmerratios} shows that when the \hab\ ratio is measured using G102 for 
H$\beta$ and G141 for H$\alpha$, the result is not dependent on roll angle, and the 
uncertainties are low: the value measured from the roll-combined spectrum is 
\hab $ = 2.96 \pm 0.1$, which is slightly higher than the values measured from 
each roll:  $2.86 \pm 0.1$ (308\arcdeg\ roll) and  $2.86 \pm 0.1$ (139\arcdeg\ roll).

By contrast, when both H$\alpha$ and H$\beta$ are measured from G141, the measured values are
higher (3.1 to 3.7), with larger uncertainties ($\pm 0.2$), and with a 
a roll dependence that is larger than the uncertainties. 

Thus, the more reliably measured Balmer ratio is \hab $ = 2.96 \pm 0.1$, with 
H$\beta$ measured from G102 and H$\alpha$ from G141, from the roll-combined spectra.
This is quite close to the intrinsic value, and indicates that the reddening is very low: 
\ebv $= 0.028 \pm 0.04$ mag.  

Higher-order Balmer line ratios are also measured:  H$\beta$/H$\gamma$ and H$\beta$/H$\delta$.
The fluxes of the higher-order Balmer line fluxes are measured less precisely than H$\beta$, 
due to their relative faintness and the blending of H$\gamma$ with [O III]~4363~\AA\ 
in the G102 grism.  Nevertheless, the reddening inferred from the higher-order Balmer line ratios
is consistent, with the uncertainties, with the more precise measurement from \hab .

\subsection{[N II] / H$\alpha$ ratio}
\begin{figure}
\includegraphics[width=3in,angle=0]{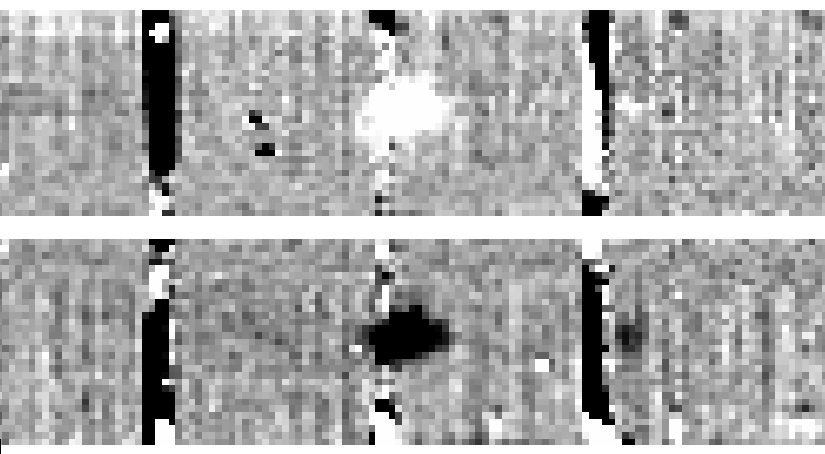}  
\includegraphics[width=3in,angle=0]{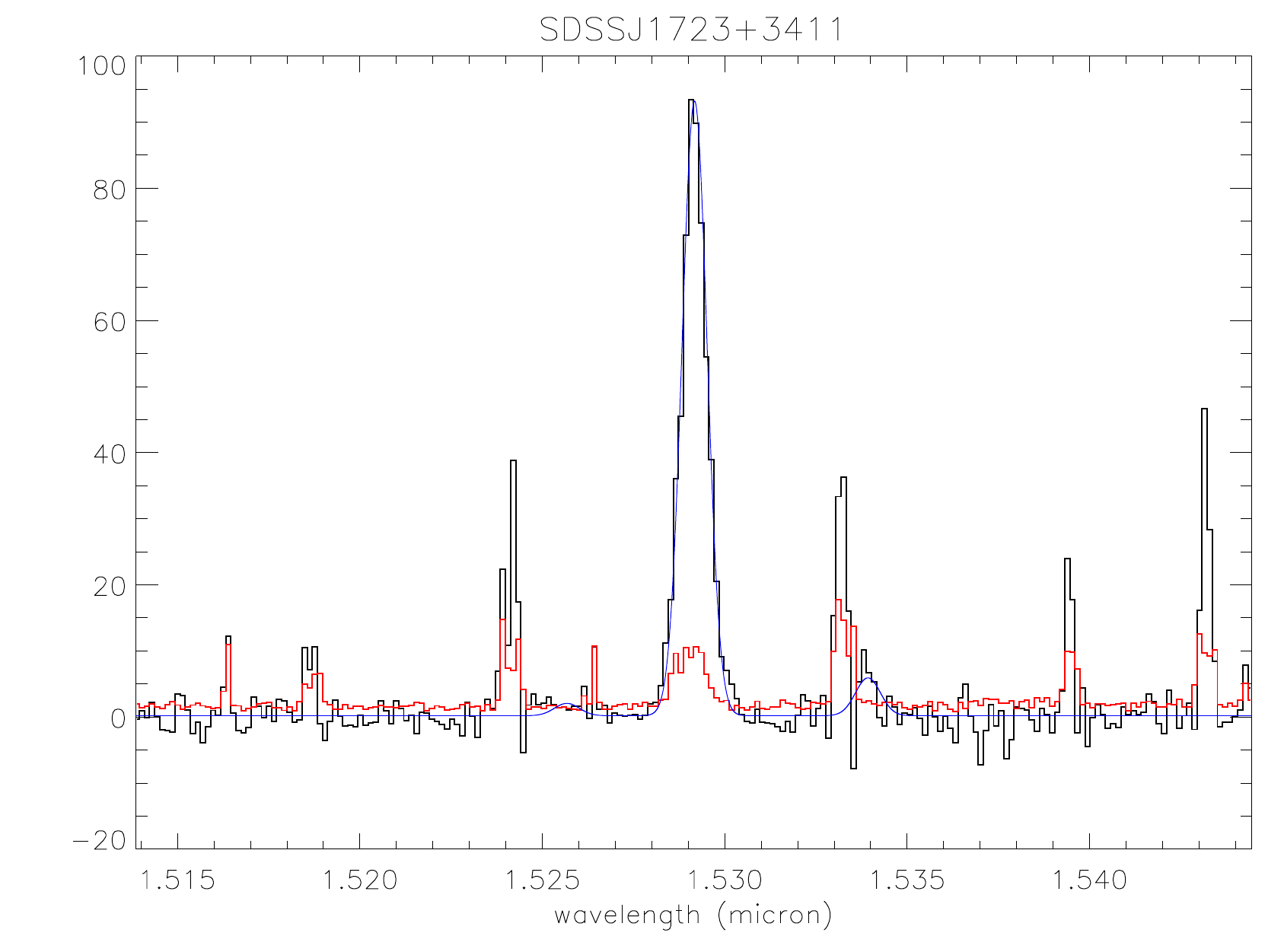}  
\figcaption{The Gemini GNIRS spectrum of \arcname .
 Left panel: The H$\alpha$ region of the two-dimensional spectrum, 
with the sum of the A nods at top, and the sum of the B nods at bottom. 
Wavelength increases to the right; H$\alpha$ is centered.  
This is a zoom in covering wavelengths of approximately 1.522--1.538~\micron.
The blue wing of H alpha is mildly contaminated by a sky line; 
[NII 6586]  falls just redward of a skyline,  and is confidently detected in both nods. 
Right panel:   The resulting extracted 1D spectrum of H$\alpha$ and [N II] in  \arcname .
Plotted is the average spectrum out of 11 integrations (black steps), 
the error in that mean (red steps), and the best-fitting Gaussians from MPFIT (blue steps).
\label{fig:gnirs_Ha}}
\end{figure}

Figure~\ref{fig:gnirs_Ha} shows the 2D GNIRS spectra, the  
extracted 1D GNIRS spectrum, and the best fit to  H$\alpha$ and [N~II].
From fitting the 1D Gemini GNIRS spectrum, we measure the 
[N~II]~6586 / H$\alpha$ flux ratio as $0.062  \pm  0.01$.  
This corresponds to an oxygen abundance
of $12 + \log(O/H) = 8.21 \pm 0.04$ \added{($31\%$ of solar)} using the linear calibration of \citet{Pettini:2004bq}, 
and $8.19 \pm 0.03$ \added{($29.5\%$ of solar)} using their third-order calibration. 
The calibration of \citet{Kewley:2019kf} yields a higher metallicity, $12 + \log(O/H) = 8.62 \pm 0.01$ 
assuming a pressure of $\log (P/k) = 6$.
As noted above, the GNIRS spectrum covered only one sub-region of giant arc, and is therefore not
strictly comparable to the measurements from Keck/ESI, MMT/Blue Channel, and \hst / WFC3-IR,  which covered
the entire giant arc.

\added{\subsection{Comparison of rest-frame optical diagnostic diagrams and equivalent widths}\label{sec:BPT}
Line ratio versus line ratio diagrams, including 
[O III]/H$\beta$ versus [N~II]/H$\alpha$ or versus [S~II]/H$\alpha$ (the so-called BPT diagrams, \citealt{Baldwin:1981ev}),
and O32 versus R23 or versus O3N2, are a standard tool to classify the spectra of galaxies.  
Multiplexed spectroscopic surveys of field galaxies at $z\sim 2$ have revealed marked evolution in these 
diagrams since $z=0$ and $z\sim2$ \citep{Kewley:2013cs, Steidel:2014es, Shapley:2015bd}.
It is therefore instructive to locate \arcname\ on these diagrams.
Figure 9 of \citet{Sanders:2016ea} plots O32 versus R23 for the SDSS sample at $z\sim0$ and the MOSDEF
sample at $z\sim2$.  Compared to both samples, \arcname\ sits at the top of both the O32--R23
relation and the O32--O3N2 relation; thus its emission lines ratios are extreme compared to both SDSS and to MOSDEF.
Green pea galaxies also inhabit the upper tip of the O32--R23 relation \citep{Jiang:2019hi}. 
Figure 11  of \citet{Sanders:2016ea} plots the [N~II] and [S~II] BPT diagrams for SDSS and for MOSDEF;
\arcname\ sits on the upper left wing of both diagrams, with emission lines that are extreme even compared
to the typical $z\sim2$ galaxies of MOSDEF.
\citet{Kewley:2013ht} argue that that regime of the BPT diagrams should be populated given
some combination of higher ionization parameters, harder ionizing spectra, and higher electron
density compared to SDSS at $z \sim 0$.
}

\added{Also striking are the high equivalent widths of the rest-frame optical emission lines in \arcname.
To be quantitative, \arcname\ has a combined rest-frame equivalent width of 
H$\beta$, [O~III]~4959, and [O~III]~5007~\AA\ of  $1029 \pm 7$\AA.  
This is even higher than the median equivalent widths of 
$670^{+260}_{-170}$~\AA, $649^{+49}_{-52}$~\AA, and $692^{+102}_{-103}$~\AA\ 
inferred through fitting photometry to color-selected $z\sim7$--8 galaxies 
of, respectively, \citet{Labbe:2013kx, deBarros:2019hy}, and \citet{Endsley:2020uu}.
This is also the regime of high equivalent width that is characteristic of the 
so-called ``green pea'' galaxies at $z\sim0.1$--$0.4$, which are selected to have 
[O~III]~5007~\AA\ equivalent widths exceeding 300~\AA\ \citep{Cardamone:2009in, Jiang:2018ub}.
}

\added{\subsection{Comparison of UV emission-line equivalent widths}\label{sec:UVEW}
Recent surveys have captured the rest-frame UV emission lines in a number of galaxies in
the local universe, and at moderate to high redshift.  The total [C~III] rest-frame equivalent width 
of \arcname, $-3.5 \pm 0.3$~\AA\ from the MMT/BC spectrum and $-3.0 \pm 0.1$ from the Keck/ESI spectrum
is within the large range observed for UV-bright $z\sim0$ galaxies \citep{Rigby:2015jy},
it matches the median equivalent width of $-3.3$~\AA\ for a sample of $z \sim 0.2$ Green Pea galaxies
\citep{Ravindranath:2020cy}, and is not quite as strong as the median equivalent width of $-4.8$~\AA\
for a sample of extreme emission line galaxies \citep{Senchyna:2020tj}.

Comparing to more distant samples, the [C~III] equivalent width is 
larger than average but not unusual for lensed $z>1$ galaxies \citep{Rigby:2015jy}.
Comparing the [C~III], He~II~1640~\AA, O~III]~1666~\AA, and [Si III]~1883~\AA\ rest-frame equivalent widths
to stacks of $z\sim 3$ Lyman alpha emitters in \citet{Feltre:2020wr}, 
the equivalent widths for \arcname\ are entirely in line with the Feltre sub-samples that are 
UV-bright, are in the same stellar mass bin, and the same bin of star formation rate.
In addition, a sample of extreme [O~III]~5007~\AA\ emitters at $1.3<z<3.7$ has a median 
rest-frame C~III] equivalent width of $-5.7$, which is about twice as high as for \arcname\ 
(from Table 5 of \citealt{Tang:2020vy}).  \oiiiuv\ was not detected in enough of their spectra to 
compare the equivalent width to \arcname.

Thus, the equivalent widths of the rest-frame UV lines in \arcname\ are entirely within the range
observed for extreme galaxies at $z \sim 0$, and at $z \sim 1$--3.  
}

\subsection{Strong emission-line diagnostics}\label{sec:seld}
\begin{figure*}
\includegraphics[width=7in,angle=0]{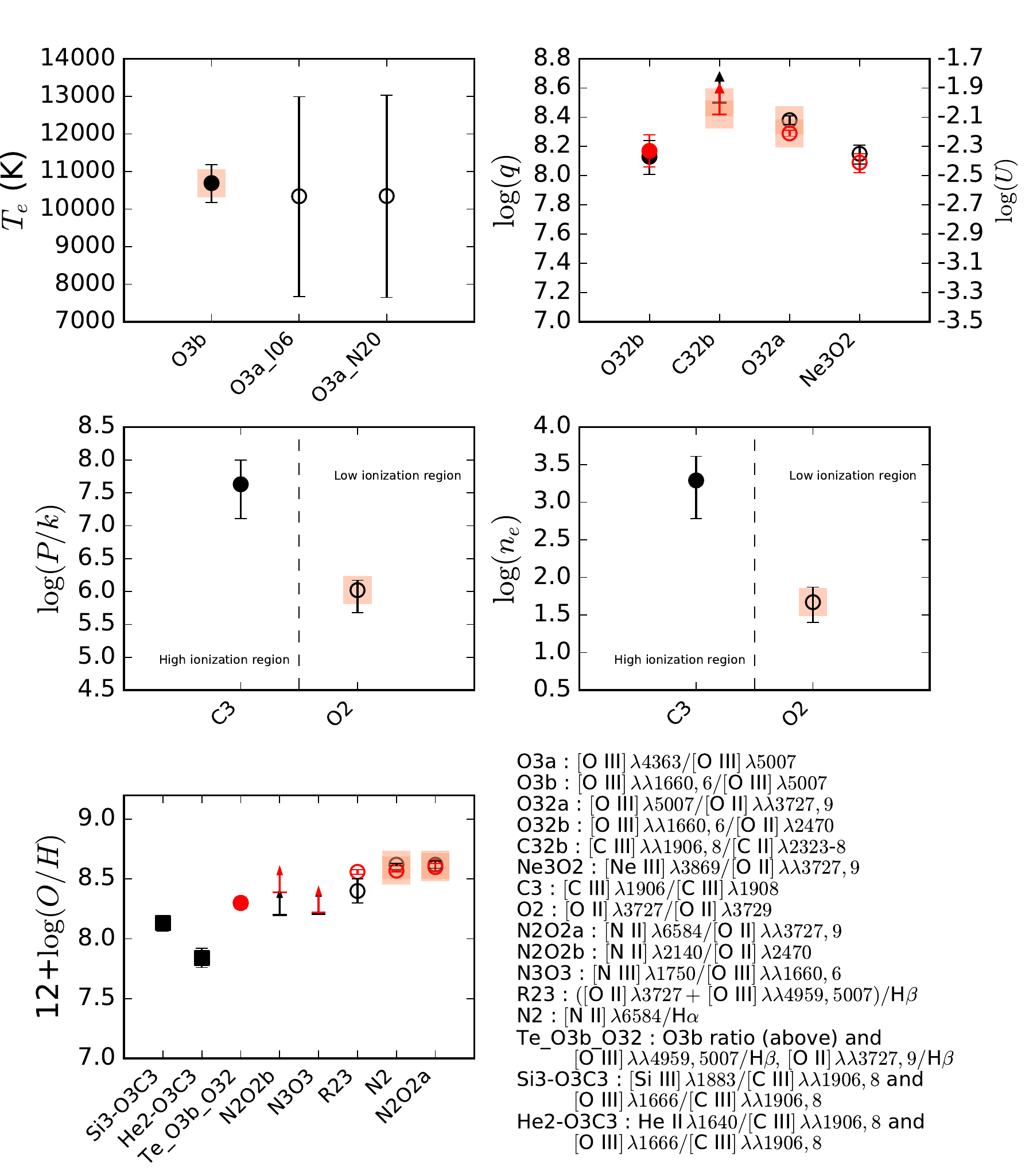}  
\figcaption{Comparison of strong emission line diagnostics for physical parameters 
of the ionized gas in \arcname .
In each panel, a point denotes the median of the  probability density function 
of the measured physical quantity generated by performing  every diagnostic 10$^4$ times; 
error bars show the 16$^{th}$ and 84$^{th}$ percentile values. 
Rest-frame UV diagnostics are denoted by filled symbols, and optical diagnostics by open symbols. 
The diagnostics we consider most reliable are highlighted in peach.
In the bottom left and top right panels, the color coding indicates the pressure 
assumed for the diagnostic: black symbols assume 
$\log (P/k) = 6$~K/cm$^{-3}$   (as inferred from the optical), and red symbols assume 
$\log (P/k) = 7$~K/cm$^{-3}$  (as inferred from the UV).  
In each panel, the x-axis lists a short-hand name for each spectral diagnostic;
the key to the labels is given at bottom right.
The spectral diagnostics are
as defined in \citet{Kewley:2019ki} and \citet{Kewley:2019kf}, and are also listed in Table~\ref{tab:diagnostics}. 
The ``\_I06'' and ``\_N20'' suffixes on the O3a $T_e$ diagnostics refer to the calibrations of that
diagnostic from \citet{Izotov:2006cd} and \citet{Nicholls:2020iy}, respectively.
The y axes of the top right panel show the ionization parameter in both its
dimensionless and dimensional forms, $U$ and $q$, where $U = q/c$.
\label{fig:seld}}
\end{figure*}

A19 intercompared the various strong emission-line diagnostics of nebular
physical conditions within a star-forming region of the $z=1.70$ lensed galaxy \rcsohthree.  
Their Table~4 describes the line ratio(s) used for each diagnostic. 
Here, we calculate the same diagnostics for \arcname, and analyze the results and 
the degree to which the diagnostics agree.
Figure~\ref{fig:seld} compares multiple strong emission line diagnostics 
for the physical parameters of the ionized gas in \arcname . 
Table~\ref{tab:inferredvalues} tabulates the inferred physical parameters.
We use diagnostics from \citet{Kewley:2019ki}, \citet{Kewley:2019kf}, 
\added{and \citet{Byler:2020gy}}; 
we do not use historical diagnostics since they used older atomic data.
We now consider each physical parameter in turn, starting at the top left of 
Figure~\ref{fig:seld}.  For each parameter, we explain which diagnostics we consider
most trustworthy and why; these are highlighted in Figure~\ref{fig:seld}.  We then compare to the
other available diagnostics, to see which diagnostics are reliable, and which are not.

{\bf Electron temperature:}
For \arcname, as was true for \rcsohthree, the best constraint on electron temperature $T_e$ 
comes from the detection of the UV auroral lines \oiiiuv , 
in combination with the rest-frame optical [O~III]~5007~\AA\ line; 
this diagnostic is marked ``O3b'' in Figure~\ref{fig:seld}.
In the MMT spectrum, only the 1666~\AA\ line is detected; the 1660~\AA\ line flux is 
inferred since the line ratio is set by atomic physics. 
The inferred $T_e$ is $10700 \pm 500$ K.  
The main drawback of this diagnostic is its susceptibility to reddening given the large 
wavelength span between 1660~\AA\ and 5007~\AA; for \arcname\ the resulting constraint is
particularly tight because the reddening is very low.

By contrast, the diagnostics which rely on the  [O~III] 4363~\AA\ auroral line 
(O3a\_I06 and O3a\_N20 in Figure~\ref{fig:seld}) provide poorer constraints
for both \arcname\ and \rcsohthree.  For \arcname\ the constraint 
is particularly poor because at the low spectral resolution of the G102 grism,  
[O~III] 4363~\AA\ is badly blended with H$\gamma$. 
This faint auroral line will be much easier to capture with \jwst .

{\bf Pressure:} 
For \arcname, the best constraints on nebular pressure come from the \oiidoublet\ ratio; 
the calibrations in \citet{Kewley:2019ki} yield an ISM pressure of 
$\log(P/k) = 6.02^{+0.15}_{-0.34}$, 
which is independent of ionization parameter.

By contrast, the ISM pressure that one would infer from the UV \ciiidoublet\ lines 
is higher by 1.5 dex, at $\log (P/k) = 7.6^{+0.37}_{-0.52}$. 
This effect is predicted to occur as a result of the UV lines arising 
in more highly ionized region of the nebula, closer to the ionizing sources, than where the [O~II]
is emitted  \citep{Kewley:2019ki}.  A similar (1 dex) offset was seen in \rcsohthree\ by A19.

{\bf Metallicity:} 
 The [N~II]~$\lambda 6584$/[O~II]~$\lambda 3727$ line ratio is the most sensitive to metallicity 
and least sensitive to the effects of ionization parameter and nebular pressure.  
It is particularly well-suited to \arcname\ since the reddening is low.  
The dereddened $\log$ [NII]/[OII] ratio of $-0.87 \pm 0.04$  
yields a metallicity of $\log(O/H)+12=8.62 \pm 0.03$,  
\added{which is $79\%$ of solar.}
 This estimate is independent of ionization parameter unless the ionization parameter is small 
(i.e. $\log(U) \lesssim -3.0$), and is also independent of ISM pressure.  
This metallicity estimate agrees with the [N~II]/H$\alpha$ \added{(''N2'')} metallicity estimate of 
$\log(O/H) + 12 = 8.62 \pm 0.01$.  
The optical R23\footnote{R23 $\equiv$ (  [O III] 4959, 5007~\AA\  + [O II] 3727, 3729~\AA\ ) / H$\beta$} 
ratio is 0.93, which is above the theoretical models used to calibrate R23.  
Large R23 values can occur when the metallicity is around the R23 local maximum, which occurs 
around $\log(O/H) + 12 = 8.4 \pm 0.1$ \added{($48\%$ of solar)}.
\added{\citet{Patricio:2018kv} noted that 10 of the 16 lensed $z\sim2$ galaxies in their
sample had R23 values exceeding the theoretical maximum.}

\added{Disappointingly, } we find that the metallicity diagnostics that use only rest-frame UV emission lines 
\added{provide very different constraints than the optical metallicity diagnostics.  Neither}
``N2O2b'' $\equiv$ [N~II]~2140 / [O II]~2470  \replaced{and}{nor} ``N3O3'' $\equiv$ [N~III]~1750 / [O~III]~1660,6
are unable to constrain the metallicity, even though the rest-frame UV spectrum is of high quality.
\added{The UV--only ``Si3-O3C3' and ``He2-O3C3''diagnostics of \citet{Byler:2020gy}\footnote{We use the
fourth-order polynomial fits published in their \S5.4.} indicate metallicities that are 0.5 and 0.8 dex 
lower, respectively, than what the optical diagnostics yield.  
This discrepancy is much larger than the uncertainties.}

The $T_e$--based diagnostic of [O~III]~1660,6 / [O~III]~5007 is able to constrain the metallicity, 
giving a value 0.3~dex below that from the optical metallicity diagnostics of 
[N~II]/[O~II] and [N~II]/H$\alpha$.
This discrepancy between the strong line diagnostics and the $T_e$ method is well known
\citep{Stasinska:2005bx, LopezSanchez:2012df}. 

{\bf Ionization parameter:}  
The most reliable ionization parameter diagnostic in this case is the optical 
``O32a'' ratio,\footnote{O32 $\equiv$ [O III] 5007~\AA\ / [O II] 3727, 3729~\AA} 
which is dependent on the ISM pressure and on the metallicity.  
Using a pressure of $\log(P/k)=6$, we derive an ionization parameter of  
$\log(U) = -2.1 \pm 0.03$, which is  $\log(q \mathrm{(cm/s)}) = 8.38 \pm 0.03$.
This value is broadly consistent with the ionization parameter that we derive from 
the UV C32b ratio of  
$\log(U) \ge -2.0$, which is  $\log(q \mathrm{(cm/s)}) \ge 8.5$.

The optical O32 diagnostic has an analogue in the UV, 
which uses the [O III]~1660,1666 / [O II]~2470 ratio 
(``O32b'' in Figure~\ref{fig:seld}.)   \citet{Kewley:2019kf} found that this
diagnostic depends strongly on metallicity, and should not be used at large 
pressures and low metallicities.   As discussed above, the pressure in the [O~III] zone, 
as inferred from the UV \ciiidoublet\ lines, is indeed extremely high.  As such,
for \arcname, the UV O32b diagnostic is less sensitive to ionization parameter than to
metallicity and pressure, and should not be used. 
Indeed, we find that the ionization parameter one would infer from the 
UV O32 diagnostic (``O32b'') is 0.25 dex lower than from optical O32 ratio (``O32a'').
A similar (0.2 dex) offset was seen by A19, with the opposite sign.  

{\bf Density:} 
If one assumes a constant electron temperature, then 
the electron density calculated using the \oiidoublet\ doublet ratio is 
$n_{e}\sim 40 \pm 10$~\cc, where the errors are due to the residual 
metallicity dependence of the [O~II] electron density relation (see \citealt{Kewley:2019ki}).
If the electron temperature is not constant, the ISM pressure is a more realistic 
quantity to measure using the [O~II] doublet; pressure and density are simply related as $n_e = P/T_e k$ .  
For the same reason as the pressure diagnostics, the UV \ciiidoublet\ diagnostic yields a much higher
density than does the optical [O~II]/[O~II] ratio.

\section{Discussion and Conclusions}			\label{sec:discussion}
This paper presents overlapping, complete spectra from Keck, MMT, Gemini, and \hst\  
that fully cover the wavelength range 
$1375 < \lambda_{rest} < 7230$~\AA\ for the extremely bright, gravitationally--lensed
galaxy \arcname\ at redshift \bestz .  The spectra have been relatively flux calibrated using bright emission lines, 
removing the possible systematic effect of relative flux errors that appear in previous work.  As such,
the spectra published electronically in this paper represent one of the highest-quality 
empirical templates of star-forming galaxies available at any redshift.  We expect that this template will 
prove useful to the community in planning observations with telescopes including \jwst .

\added{\arcname\ has an intrinsic UV luminosity that is very close to what is characteristic
for star-forming galaxies at its redshift. Its star formation rate is a factor of 12 above the 
star formation rate--stellar mass relation at its redshift, indicating that it is a ``starburst'' 
galaxy experiencing much more rapid star formation than is typical for galaxies of its stellar mass
at redshift $z=1.3$. Though \arcname\ has a modest redshift of \bestz , its properties are consistent 
with the most extreme
galaxies known.  Its emission line ratios place it in the extreme tip of the O32--R23 and O32--O3N2 
diagnostic diagrams and in the upper left wing of the [N~II] and [S~II] BPT diagrams, and has the extremely 
high equivalent widths of H$\beta$ and \oiiidoublet\ that are characteristic of extreme emission line galaxies 
like the ``green peas'' at $z\sim 0.3$ and star-forming galaxies at $z\sim 7$--8x.  Its rest-frame UV 
emission line equivalent widths are within the range observed for extreme galaxies in the nearby and 
distant universe. As such, \arcname\ should serve as an appropriate template for 
extreme star-bursting galaxies at any redshift.}

We infer the nebular physical conditions within this galaxy --- parameterized as 
electron temperature, pressure, density, metallicity, and ionization parameter --- using
widely used rest-frame optical diagnostics as well as rest-frame UV diagnostics that have been 
developed for use at high redshift by \jwst , but have not been well tested.  
We find that \arcname\ has a metallicity close to solar, and a 
higher ionization parameter and higher ISM pressure than local 
star-forming galaxies of the same metallicity \citep{Kewley:2006ib, Thomas:2019gu}. 
This is similar to results for other high redshift star forming galaxies 
\citep{Hainline:2009fg, Bian:2010bl, Liu:2008iq, Nakajima:2013ir, Shirazi:2014fm, Shirazi:2014fe}, 
and supports a picture in which the mode of star formation at cosmic noon was quite different than 
how stars form in the local universe (e.g.\ \citealt{Rigby:2008fw, Rujopakarn:2012hc}).

We measure dramatically higher (1.5 dex) pressure in \arcname\ from the rest-frame UV
\ciiidoublet\ diagnostic compared to the \oiidoublet\ doublet ratio.
A19 found the same offset in a lensed galaxy at $z=1.70$.  
While to be expected given that \ciiidoublet\ arises from hotter nebular regions than the
 \oiidoublet\ \citep{Kewley:2019ki}, these results should be a 
sobering caution against inter-comparing pressures derived from the UV versus the optical diagnostics.  
If care is not taken, high-redshift galaxies may look more extreme than local 
galaxies \emph{simply because of the diagnostics used.}  

\added{We find that the rest-frame optical strong-line diagnostics N2O2a, N2, and R23 provide
consistent metallicity estimates.  This is consistent with the good performance of R23 and N2 
seen by \citet{Patricio:2018kv} when comparing to the [O~III]~4363~\AA\ auroral line method;
they did not examine N2O2a.  In \arcname, we do not see an offset in the metallicity diagnostics 
that use N2, as was seen by  \citet{Patricio:2018kv}.}

Unfortunately, we find that no currently--used diagnostic that uses only rest-frame UV emission 
lines is able to \replaced{meaningfully constrain the metallicity.}{successfully determine the
metallicity, as judged against the rest-frame optical diagnostics (N2, N2O2a, and R23).}
This is consistent with the results of A19 for a different lensed galaxy at $z=1.7$.  
The metallicity is effectively constrained
for \arcname\ using two diagnostics that incorporate [N~II]~6584~\AA , as well as 
via the electron temperature method that uses  \oiiiuv\ to [O~III]~5007~\AA , though 
with the expected offset between the electron temperature and strong-line methods.  

We now consider what these results mean for \jwst\ studies of galaxies at high 
redshift, especially at the epoch of reionization. The NIRSpec instrument's 
wavelength range of 0.7--5~\micron\ can capture the following key emission lines 
for galaxies in a multiplexed way out to the following redshifts:
\begin{itemize}
\item \oiiiuv\          for $z>3.2$, 
\item \ciiidoublet\     for $z>2.7$, 
\item \oiidoublet\      for $0.88 < z < 12.4$, 
\item \oiiidoublet\     for $0.40 < z <  9.0$, and 
\item  H$\alpha$ plus [N~II]~6584~\AA\ for $0.06 < z <  6.6$.  
\end{itemize}

The metallicity diagnostics that work best for \arcname , which use the optical [N~II] lines,  
will for galaxies at the reionization epoch be redshifted out of the \jwst\ NIRSpec wavelength range.
Indeed, this reality has fueled interest in UV--only metallicity diagnostics that would 
be well-suited for \jwst\ studies of galaxies at the epoch of reionization. 
Unfortunately, given the \replaced{inability of the }{failure of all the } 
UV--only metallicity diagnostics 
\replaced{[N~II]~2140/[O~II]~2470~\AA\ (``N2O2b'') and [N~III]~1750/[O~III]~1660,6~\AA\ (``N3O3'') to constrain}
{(N2O2b, N3O3, Si3-O3C3, and He2-O3C3) to correctly determine }
the metallicity of \arcname\ in this study, and similar results by A19, 
we are pessimistic about the effectiveness of these diagnostics at moderate or at high redshift.
It would be timely to develop alternative rest-frame UV metallicity diagnostics, based on 
rest-frame UV spectral atlases at moderate redshift (for example \megasaura ; \citealt{Rigby:2018ev})
and at low redshift, like the upcoming ULYSSES and CLASSY programs with \hst .

In contrast to the failure of the rest-frame UV metallicity diagnostics, 
the electron temperature method of measuring metallicity, 
using the ratio of [O~III]~1660,6~\AA\ or [O~III]~4363~\AA\ 
to [O~III]~5007~\AA , has proven effective at $z\sim 1$--2 (this work, A19, and \citealt{Steidel:2016jv}),
and will be available to \jwst\ for galaxies with redshifts below $z=9$.  
Therefore, we suggest that 
these electron temperature diagnostics are particularly well-suited for \jwst, with the caveat 
that there is a well-known offset between the electron temperature methods and the strong emission 
line diagnostics.  We therefore suggest that \jwst\ spectroscopic surveys should be sure to cover
either or both of these two auroral emission lines.  
In addition, we suggest that \jwst\ spectroscopic surveys of the epoch of reionization should invest the 
exposure time to capture the rest-frame optical emission lines at the red edge of NIRSpec's bandpass.
In particular, based on the results of this work, we argue that NIRSpec surveys of $z \le 9$ galaxies 
should capture both the \oiidoublet\ and \oiiidoublet\ doublets.  Even though this strategy requires 
the serial use of two gratings rather than one, we believe it is necessary to obtain reliable
measurements of metallicity, pressure, and ionization parameter.
As an example, for $z=9$, the NIRSpec G235M/F170LP grating/filter pairing captures \oiiiuv\ and \ciiidoublet, 
while G395M/F290LP captures [O~II]~3727, [O~III]~4363, and [O~III]~5007~\AA .
It may also make sense to obtain MIRI spectra for a subset of the survey
to capture H~$\alpha$ and  [N~II]~6584~\AA , even though this must be one galaxy at a time.
For galaxies with redshift $z>9$,  while the optical [O~III] lines will not be accessible to
\jwst / NIRSpec, it would still be worth obtaining the optical [O~II] lines to measure the 
ISM pressure.

\acknowledgments
Acknowledgments:  
We thank Ramesh Mainali for commenting on the draft manuscript.
We thank Glenn Kacprzack for assistance reducing the ESI/Keck spectra.
Based on observations made with the NASA/ESA Hubble Space Telescope,
obtained from the Data Archive at the Space Telescope Science
Institute, which is operated by the Association of Universities for
Research in Astronomy, Inc., under NASA contract NAS 5-26555. These
observations are associated with program \# 14230.
Support for program 14230 was provided by NASA through a grant from
the Space Telescope Science Institute, which is operated by the
Association of Universities for Research in Astronomy, Inc., under
NASA contract NAS 5-26555.
Some of the data presented herein were obtained at the W.M. Keck
Observatory, which is operated as a scientific partnership among the
California Institute of Technology, the University of California and
the National Aeronautics and Space Administration. The Observatory was
made possible by the generous financial support of the W.M. Keck
Foundation.  We acknowledge the very significant cultural role and 
reverence that the summit of Maunakea has always had within the 
indigenous Hawaiian community. We are most fortunate to have 
the opportunity to conduct observations from this sacred mountain.
Some of the observations reported here were obtained at the MMT Observatory, 
a joint facility of the University of Arizona and the Smithsonian Institution. 
Some of the data presented herein were obtained at the Gemini
Observatory, which is operated by the Association of Universities for
Research in Astronomy, Inc., under a cooperative agreement with the
NSF on behalf of the Gemini partnership: the National Science
Foundation (United States), the National Research Council (Canada),
CONICYT (Chile), Ministerio de Ciencia, Tecnología e Innovación
Productiva (Argentina), and Ministério da Ciência, Tecnologia e
Inovação (Brazil).

\bibliographystyle{astroads}
\bibliography{papers}

\startlongtable
\begin{deluxetable}{llllllllll}
\tablecolumns{10}
\tablewidth{0pc}
\tablecaption{Fluxes for nebular emission lines in \arcname. \label{tab:linefluxes}}
\tablehead{ \colhead{Line ID} & \colhead{$\lambda_{\mathrm{rest}}$} & \colhead{telescope/spectrograph} & \colhead{W$_{\mathrm{r,fit}}$} & \colhead{$\Delta$ W$_{\mathrm{r,fit}}$}  & \colhead{W$_{\mathrm{r,signi}}$} & \colhead{flux}               & \colhead{$\Delta$ flux}     & \colhead{flux$_{\mathrm{dr}}$} & \colhead{$\Delta$ flux$_{\mathrm{dr}}$}}
%
\startdata
                 O I 1304 &                   1304.86 &          MMT/BC &          $>-$5.0  &                      - &                      - &  $<-$4.9 &             - &                     $<-$6.0 &             - \\
                 O I 1306 &                   1306.03 &          MMT/BC &          $>-$5.0  &                      - &                      - &  $<-$4.7 &             - &                     $<-$5.8 &             - \\
               Si II 1309 &                   1309.28 &          MMT/BC &          $>-$5.7  &                      - &                      - &  $<-$4.6 &             - &                     $<-$5.8 &             - \\
               C II 1335a &                   1334.58 &          MMT/BC &          $>-$12   &                      - &                      - &   $<$4.1 &             - &                      $<$5.1 &             - \\
               C II 1335b &                   1335.66 &          MMT/BC &          $>-$11   &                      - &                      - &   $<$4.05 &             - &                      $<$5.0 &             - \\
               C II 1335c &                   1335.71 &          MMT/BC &          $>-$11   &                      - &                      - &   $<$4.05 &             - &                      $<$5.0 &             - \\
               He II 1640 &                   1640.42 &          MMT/BC &          $-$0.64  &                   0.1  &                   11. &      6.4 &          1.4 &                         7.8 &                1.8 \\
              O III] 1660 &                   1660.81 &          MMT/BC &         $>-$0.15  &                      - &                      - &      1.7\tablenotemark{a} &         0.4\tablenotemark{a} &                         2.1 &               0.5 \\
              O III] 1666 &                   1666.15 &          MMT/BC &          $-$0.48  &                   0.1  &                   9.5 &      5.0  &         1 &                         6.1 &                1.5 \\
              N III] 1750 &                    1749.7 &          Keck/ESI &         $>-$0.21  &                      - &                      - &   $<$1.0 &             - &                      $<$1.22 &               - \\
          {[}Si III] 1882 &                   1882.71 &          Keck/ESI &          $-$0.71  &                   0.08 &                   16   &      7.0 &          0.8 &                        8.6 &                1.0 \\
             Si III] 1892 &                   1892.03 &          Keck/ESI &          $-$0.36  &                   0.08 &                   8.3 &      3.5 &          0.7 &                         4.3  &                0.9 \\
             {[}C III] 1906 &                 1906.68 &          MMT/BC &           $-$2.05 &                   0.2 &                    34 &       17  &          1.4 &                        21.3 &                1.7 \\
             {[}C III] 1906 &                 1906.68 &          Keck/ESI &           $-$1.78 &                   0.08 &                    38  &     17.3  &         0.8 &                        21.3 &               0.96 \\
                C III] 1908 &                 1908.73 &          MMT/BC &           $-$1.4  &                   0.2 &                    24&       12.0  &          1 &                        14.85 &                1.6\\
              C III] 1908 &                   1908.73 &          Keck/ESI &           $-$1.23 &                   0.08 &                    27 &      12.0 &          0.8 &                        14.8  &                1 \\
               N II] 2140 &                   2139.68 &          MMT/BC &         $>-$0.228 &                      - &                      - &   $<$0.73 &             - &                      $<$0.94 &                 - \\
           {[}O III] 2320 &                   2321.66 &          Keck/ESI &         $>-$0.163 &                      - &                      - &   $<$0.47 &             - &                      $<$0.59 &                 - \\
               C II] 2323 &                   2324.21 &          Keck/ESI &         $>-$0.162 &                      - &                      - &   $<$0.46 &             - &                      $<$0.58 &                 - \\
              C II] 2325c &                   2326.11 &          Keck/ESI &          $-$0.86  &                   0.2  &                   15   &      5.7  &           1.2 &                         7.1  &                1.5 \\
              C II] 2325d &                   2327.64 &          Keck/ESI &          $-$0.22  &                   0.1  &                   3.9  &      1.4 &            0.9 &                         1.8 &                 1.1 \\
               C II] 2328 &                   2328.84 &          Keck/ESI &         $>-$0.156 &                      - &                      - &   $<$0.45 &             - &                      $<$0.55 &            - \\
             Si II] 2335a &                   2335.12 &          Keck/ESI &         $>-$0.157 &                      - &                      - &   $<$0.45 &             - &                      $<$0.55 &            - \\
             Si II] 2335b &                   2335.32 &          Keck/ESI &         $>-$0.155 &                      - &                      - &   $<$0.45 &             - &                      $<$0.55 &            - \\
               Fe II 2365 &                   2365.55 &          Keck/ESI &          $-$0.26  &                   0.07 &                   7.3  &      1.7  &           0.4 &                         2.1 &             0.5 \\
              Fe II 2396a &                   2396.15 &          Keck/ESI &           $-$2.1  &                   0.2  &                   58   &     13.7 &            1   &                        16.8 &             1  \\
              Fe II 2396b &                   2396.36 &          Keck/ESI &          $-$0.265 &                   0.1  &                   7.3  &      1.7 &            0.8 &                         2.1 &             1.0 \\
            {[}O II] 2470 &                   2471.03 &          Keck/ESI &           $-$1.19 &                   0.1  &                   26   &      7.2 &            0.7 &                         8.7 &             0.8 \\
               Fe II 2599 &                   2599.15 &          Keck/ESI &         $>-$0.089 &                      - &                      - &   $<$0.21 &             - &                      $<$0.24 &              - \\
               Fe II 2607 &                   2607.87 &          Keck/ESI &        $>-$0.1075 &                      - &                      - &   $<$0.25 &             - &                      $<$0.29 &              - \\
               Fe II 2612 &                   2612.65 &          Keck/ESI &         $-$0.41   &                   0.07 &                    11  &      2.15 &          0.4 &                         2.6  &             0.5 \\
               Fe II 2614 &                   2614.61 &          Keck/ESI &         $>-$0.11  &                      - &                      - &   $<$0.25 &             - &                      $<$0.29 &              - \\
               Fe II 2618 &                    2618.4 &          Keck/ESI &          $>-$0.11 &                      - &                      - &   $<$0.25 &             - &                      $<$0.29 &              - \\
               Fe II 2621 &                   2621.19 &          Keck/ESI &         $>-$0.11  &                      - &                      - &   $<$0.24 &             - &                      $<$0.28 &              - \\
               Fe II 2622 &                   2622.45 &          Keck/ESI &         $>-$0.11  &                      - &                      - &   $<$0.25 &             - &                      $<$0.29 &              - \\
               Fe II 2626 &                   2626.45 &          Keck/ESI &          $-$0.58  &                   0.07 &                   15   &      3.0  &          0.4 &                         3.6  &             0.4 \\
               Fe II 2629 &                   2629.08 &          Keck/ESI &         $>-$0.11  &                      - &                      - &   $<$0.24 &             - &                      $<$0.28 &                - \\
               Fe II 2631 &                   2631.83 &          Keck/ESI &        $>-$0.11   &                      - &                      - &   $<$0.24 &             - &                      $<$0.28 &                - \\
               Fe II 2632 &                   2632.11 &          Keck/ESI &          $-$0.17  &                   0.09 &                   4.7  &       0.9 &          0.5 &                         1.0  &              0.6 \\
              Mg II 2797  &                   2798.76 &          Keck/ESI &         $>-$0.18  &                      - &                      - &   $<$0.35 &             - &                      $<$0.40 &                - \\
              Mg II 2797  &                   2803.53 &          Keck/ESI &          $-$0.69  &                   0.09 &                   9.5  &      3.1  &          0.4 &                         3.7  &              0.5 \\
                He I 3187 &                   3188.67 &          Keck/ESI &          $-$1.1   &                   0.1  &                   21   &      4.15 &          0.5 &                         4.8  &              0.5 \\
              Ne III 3342 &                   3343.14 &          Keck/ESI &          $-$0.21  &                   0.07 &                   3.7  &      0.7  &          0.2 &                         0.84 &              0.3 \\
               S III 3721 &                   3722.69 &          Keck/ESI &          $-$0.64  &                   0.1  &                   8.4  &      1.8  &          0.3 &                         2.1 &               0.4 \\
            {[}O II] 3727 &                  3727.092 &          Keck/ESI &          $-$26.9  &                   0.3  &                   260 &       76  &            0.7 &                         86.2 &              0.84 \\
            {[}O II] 3727 &                   3727.09 &         HST/WFC3-IR G102 &  $-$55    &                   1  &                     - &         77  &            2 &                         87.5 &              2 \\
            {[}O II] 3729 &                  3729.900 &          Keck/ESI &          $-$37.8  &                   0.2  &                   350 &      107  &            0.7 &                         121.0  &            0.8 \\  
            {[}O II] 3729 &                   3729.88 &         HST/WFC3-IR G102 &  $-75.$   &                   2  &                     - &       106  &             2 &                          120. &              3 \\
                  H$\eta$ &                   3836.48 &          Keck/ESI &            $-2.1$ &             0.1 &                  25    &       5.5 &          0.35 &                          6.2 &               0.4 \\
          {[}Ne III] 3869 &                   3869.86 &         HST/WFC3-IR G102 &      $-33$     &                4 &                      - &     52   &          6 &                          59 &                7 \\
           H$\zeta$; He I &                   3890.15 &         HST/WFC3-IR G102 &      $-13$     &                5 &                      - &     20.  &          8 &                          23 &                8.5 \\
          {[}Ne III] 3968 &                   3968.59 &         HST/WFC3-IR G102 &      $-10$     &               1 &                      - &     17  &          2 &                         18.7 &              2 \\
              H$\epsilon$ &                    3971.2 &         HST/WFC3-IR G102 &      $-7.4$    &               2.5&                      - &     12  &          4&                         13.4 &              5 \\
         He I 4025; He II &                      4025 &         HST/WFC3-IR G102 &      $-5.1$    &               2  &                      - &      8 &          4 &                         9.2 &               4 \\
                H$\delta$ &                   4102.89 &         HST/WFC3-IR G102 &      $-16.5$   &               2  &                      - &     26.  &        4 &                         29 &                 4 \\
                H$\gamma$ &                   4341.68 &         HST/WFC3-IR G102 &      $-43$     &               3  &                      - &     64  &          5 &                         71 &                 6 \\
               O III 4363 &                   4364.44 &         HST/WFC3-IR G102 &      $-1.9$    &               4  &                      - &      2.8 &          6 &                         3.1 &                7 \\
                He I 4471 &                    4472.7 &         HST/WFC3-IR G102 &      $-3.9$    &               3  &                      - &       5.4 &         4 &                         6.0  &               4.2 \\
               He II 4685 &                   4687.02 &         HST/WFC3-IR G102 &      $-3.1$    &               3  &                      - &       4 &           4 &                         4.3  &               4.2 \\
           {[}Ar IV] 4741 &                   4741.45 &         HST/WFC3-IR G102 &      $-3.5$    &               3  &                      - &      4. &          4 &                         4.8 &                4 \\
                 H$\beta$ &                   4862.68 &         HST/WFC3-IR G102 &     $-119$     &               3  &                      - &       142 &          4 &                          156 &               5 \\
           {[}O III] 4959 &                    4960.3 &         HST/WFC3-IR G141 &     $-220$     &               2  &                      - &    258    &          4 &                          282 &               4 \\
           {[}O III] 5007 &                   5008.24 &         HST/WFC3-IR G141 &     $-690$     &               6  &                      - &    776    &          11 &                          849 &              12 \\
                He I 5875 &                   5877.59 &         HST/WFC3-IR G141 &     $-14  $    &               8&                      - &      13   &            7.5 &                         14   &               8 \\
     O I 6300; S III 6312 &                      6310 &         HST/WFC3-IR G141 &     $-4.0$     &               10 &                      - &      3.1  &          8 &                         3.3 &                8 \\
                N II 6549 &                   6549.85 &         HST/WFC3-IR G141 &     $-12.5$    &               0.2&                      - &       8.5 &          0.2 &                         9.1 &                0.2 \\
                H$\alpha$ &                   6564.61 &         HST/WFC3-IR G141 &     $-624$     &               12 &                      - &       421 &          8 &                          449 &               8  \\
            {[}N II] 6584 &                   6585.28 &         HST/WFC3-IR G141 &     $-39$      &               1  &                      - &      26.1 &          0.5 &                        27.85 &               0.5 \\
 {[}S II] 6717; S II 6731 &                 6718+6732 &         HST/WFC3-IR G141 &     $-63$      &               13 &                      - &       38  &          8 &                         40   &               9 \\
\enddata
\tablecomments{
  Columns are:
  1) line identification ;
  2) rest-frame vacuum wavelength (\AA );
  3) which telescope and spectrograph was used to make the measurement.  ``BC'' stands for the Blue Channel spectrograph.  For \hst\ we also list which 
  4) W$_{\mathrm{r,fit}}$, the best-fit rest-frame equivalent width (in \AA ), with the
sign convention that negative equivalent width indicates emission, and positive indicates absorption.  
  5) corresponding uncertainty on 4);
  6) significance of the emission line detection;
  7) observed emission line flux, in units of 10$^{-17}$~\cgsflux ;
  8) corresponding uncertainty on 7), same units;
  9) dereddened emission line flux, same units, assuming a \citet{Cardelli:1989dp} extinction law; 
  10) corresponding uncertainty on 9), same units.
For the cases of the \oiidoublet\ and the \ciiidoublet, fluxes are reported from each of the two spectrographs that observed them.
}
\tablenotetext{a}{For the \oiiiuv\ doublet,  the 1666~\AA\ line is detected in the MMT spectrum
while the 1660~\AA\ line is not.  Since the line ratio is set by atomic physics, we infer a 
line flux for the 1660~\AA\ line based on the flux of the 1666~\AA\ line.}
\end{deluxetable}
\clearpage 

\begin{deluxetable*}{llllcc}
\tablecolumns{6}
\tablewidth{0pc}
\tablecaption{Measured Balmer line ratios and inferred \ebv\ reddening for \arcname . \label{tab:balmerratios}}
\tablehead{\colhead{line ratio} & \colhead{grism(s)} & \colhead{value} & \colhead{$\sigma$} & \colhead{inferred E(B-V)} &  \colhead{$\sigma$}}
\startdata
\cutinhead{both rolls}
 H$\alpha$/H$\beta$ &  G141/G102 &  2.96  &   0.1     &  0.028 &     0.04   \\
 H$\alpha$/H$\beta$ &       G141 &  3.73 &    0.2     &  0.26  &     0.06   \\
 H$\beta$/H$\gamma$ &       G102 &  2.24 &    0.2     &  0.09  &     0.2   \\
 H$\beta$/H$\delta$ &       G102 &  5.48 &    0.8     &  0.45  &     0.2   \\
\cutinhead{308\arcdeg\ roll}                           
 H$\alpha$/H$\beta$ &  G141/G102 &  2.86 &    0.1     & -0.007 &     0.04   \\
 H$\alpha$/H$\beta$ &       G141 &  3.37 &    0.2     &  0.16  &     0.07   \\
 H$\beta$/H$\gamma$ &       G102 &  2.36 &    0.3     &  0.2   &     0.2    \\
 H$\beta$/H$\delta$ &       G102 &  5.16 &    0.8     &  0.4   &     0.2   \\
\cutinhead{139\arcdeg\ roll}                           
 H$\alpha$/H$\beta$ &  G141/G102 &  2.86 &    0.1     & -0.006 &     0.05   \\
 H$\alpha$/H$\beta$ &       G141 &  3.11 &    0.2     &  0.08  &     0.06   \\
 H$\beta$/H$\gamma$ &       G102 &  1.98 &    0.2     & -0.15  &     0.2    \\
 H$\beta$/H$\delta$ &       G102 &  5.18 &    1.0     &  0.4   &     0.25   \\
\enddata
\tablecomments{
We report values from the summed grism spectra as well as from summed spectra from each roll angle.  
In most cases, both lines in a ratio were measured in a single grism, indicated in the ``grism(s)'' column;
an exception is H$\alpha$/H$\beta$, where the more reliable measurement of the ratio takes the  H$\alpha$
flux from the G141 grism and the H$\beta$ flux from the G102 grism.
Uncertainties are given in the columns marked $\sigma$.  
The intrinsic ratios used to calculate the reddening are from an isobaric, MAPPINGS-V photoionization \
model that assumes spherical geometry and $\log{(\mathrm P/k)}=6$, $\log{(\mathrm q)}=8$ and $\log{(\mathrm O/H)} + 12=8.53$ .}
\end{deluxetable*}

\clearpage 

\startlongtable
\begin{deluxetable*}{lll}
\tablecolumns{3}
\newcommand{\ciiiblue}{[\ion{C}{3}]$\lambda$1906}
\newcommand{\ciiired}{\ion{C}{3}]$\lambda$1908}
\newcommand{\ciiisum}{[\ion{C}{3}]$\lambda$1906,8}

\tablewidth{0pc}
\tablecaption{Description of emission-line diagnostics used, for each set of physical conditions\label{tab:diagnostics}}
\tablehead{\colhead{Diagnostic}  & \colhead{Line ratio used}  & \colhead{Reference}}
\startdata
\cutinhead{Electron temperature \Te }
Te\_O3b        & [\ion{O}{3}]$\lambda $5007/\ion{O}{3}]$\lambda\lambda$ 1660,6 & \citet{Nicholls:2020iy} \\
Te\_O3a\_I06   & [\ion{O}{3}]~$\lambda \lambda$4959,5007 $/$  [\ion{O}{3}]~$\lambda$4363 & \citet{Izotov:2006cd} \\
Te\_O3a\_N20   & [\ion{O}{3}]$\lambda \lambda$5007/[\ion{O}{3}]$\lambda$4363 & \citet{Nicholls:2020iy}  \\
\cutinhead{Density \Ne\ or Pressure \lpok}
C3             & [\ciiired/\ciiiblue & \citet{Kewley:2019ki} \\
O2             & [\ion{O}{2}]$\lambda$3729/[\ion{O}{2}]$\lambda$3727 & " \\
\cutinhead{Metallicity \logOH}
Te\_O3b\_O32   & Te\_O3b ratios (as above), [\ion{O}{3}]$\lambda\lambda$4959,5007/H$\beta$, and &  \citet{Izotov:2006cd} \\
\hspace{3mm}(direct method)& \hspace{3mm}[\ion{O}{2}]$\lambda\lambda$3727,9/H$\beta$ &  \\
N2O2b          & [\ion{N}{2}]$\lambda$2140/[\ion{O}{2}]$\lambda$2470 & \citet{Kewley:2019kf} \\
N3O3           & [\ion{N}{3}]$\lambda$1750/[\ion{O}{3}]$\lambda\lambda$1660,1666 & " \\
R23            & ([\ion{O}{2}]$\lambda$3727,9+[\ion{O}{3}]$\lambda\lambda$4959,5007)/H$\beta$ & " \\
N2             & [\ion{N}{2}]$\lambda$6584/H$\alpha$ & " \\
N2O2a          & [\ion{N}{2}]$\lambda$6584/[\ion{O}{2}]$\lambda\lambda$3727,3729 & " \\
\added{Si3-O3C3} & Si~III~1883/\ciiisum\ and  [O~III]~1666/\ciiisum    & \citet{Byler:2020gy} \\
\added{He2-O3C3} & He~II~1640/\ciiisum\  and  [O~III]~1666/\ciiisum    & \citet{Byler:2020gy} \\
\cutinhead{Ionization paramater \logq\ or $\log(U)$}
O32b           & [\ion{O}{3}]$\lambda\lambda$1660,6/[\ion{O}{2}]$\lambda$2470a,b & " \\
C32b           & [\ion{C}{3}]$\lambda\lambda$1906,8,8/[\ion{C}{2}]$\lambda$2323-8 & "  \\
O32a          & [\ion{O}{3}]$\lambda$5007/[\ion{O}{2}]$\lambda\lambda$3727,9 & " \\
Ne3O2          & [\ion{Ne}{3}]$\lambda$ 3869/[\ion{O}{2}]$\lambda\lambda$3727,9 & " \\
\enddata
\end{deluxetable*}
\clearpage 

\begin{table}
\rule{0pt}{4ex} 
\begin{tabular}{lcc|lcc}
	\toprule
	\hline
	\multicolumn{3}{c|}{UV diagnostics} & \multicolumn{3}{c}{Optical diagnostics} \\
	\hline
	Diagnostics & \multicolumn{2}{c|}{Measured values} & Diagnostics & \multicolumn{2}{c}{Measured values} \\
	\hline
	\multicolumn{6}{c}{Electron temperature,  $\times$10$^4$ K} \\
	\hline
	O3b & \multicolumn{2}{c}{1.07 $^{+0.05}_{-0.05}$} & O3a\_I06 & \multicolumn{2}{c|}{1.03 $^{+0.27}_{-0.27}$} \\
	- & \multicolumn{2}{c}{-} & O3a\_N20 & \multicolumn{2}{c|}{1.03 $^{+0.27}_{-0.27}$} \\
	\hline
	\multicolumn{6}{c}{ISM pressure, $\log P/k$  ($K/cm^3$)} \\
	\hline
	C3 & \multicolumn{2}{c}{7.63 $^{+0.37}_{-0.52}$}  & O2 & \multicolumn{2}{c|}{6.02 $^{+0.15}_{-0.34}$} \\
	\hline
	\multicolumn{6}{c}{Oxygen abundance, {\logOH}} \\
	\hline
	& For $\log{(P/k)} = 6$ & For $\log{(P/k)} = 7$ & & For $\log{(P/k)} = 6$ & For $\log{(P/k)} = 7$ \\
	\hline	
	Direct(Te\_O3b\_O32) & 8.30 $^{+0.05}_{-0.05}$ & 8.30 $^{+0.05}_{-0.05}$ & R23 & 8.4 $^{+0.1}_{-0.1}$ & 8.56 $^{+0.02}_{-0.02}$ \\
	N2O2b & $\geq$ 8.20 & $\geq$ 8.39 & N2 & 8.62 $^{+0.01}_{-0.01}$ & 8.57 $^{+0.01}_{-0.01}$ \\
	N3O3 & $\geq$ 8.21 & $\geq$ 8.22 & N2O2a & 8.62 $^{+0.03}_{-0.03}$ & 8.60 $^{+0.03}_{-0.03}$ \\
        \added{Si3-O3C3} &   \multicolumn{2}{c}{$8.13 \pm 0.07$}  & & & \\
        \added{He2-O3C3} &   \multicolumn{2}{c}{$7.84 \pm 0.08$}  & & & \\
	\hline
	\multicolumn{6}{c}{Ionization parameter, $\log q$ ($cm/s$)} \\
	\hline
	& For $\log{(P/k)} = 6$ & For $\log{(P/k)} = 7$ & For $\log{(P/k)} = 6$ & For $\log{(P/k)} = 7$ \\
	\hline
	O32b & 8.13 $^{+0.11}_{-0.12}$ & 8.17 $^{+0.11}_{-0.11}$ & O32a & 8.38 $^{+0.03}_{-0.03}$ & 8.29 $^{+0.02}_{-0.02}$ \\
	C32b & $\geq$ 8.5 & $\geq$ 8.42 & Ne3O2 & 8.15 $^{+0.06}_{-0.07}$ & 8.09 $^{+0.06}_{-0.07}$ \\
	\hline
	\multicolumn{6}{c}{Electron density, $\log n_e$ ($cm^{-3}$)} \\ 
	\hline
	C3 & \multicolumn{2}{c}{3.29 $^{+0.32}_{-0.51}$} & O2 & \multicolumn{2}{c|}{1.67 $^{+0.20}_{-0.27}$} \\
	\hline
\end{tabular}
\caption{Inferred values for physical properties, from the emission-line diagnostics.}
\label{tab:inferredvalues}

  \end{table}

\end{document}